\documentclass[11pt,preprint]{aastex}

\usepackage{lineno}
\usepackage{lscape}
\usepackage{multirow}
\usepackage{natbib}
\usepackage{times}
\usepackage{subfigure} 
\usepackage{rotating} 
 
\linenumbers
 
\def\F{{\it Fermi}-LAT } 
\def\SU{{\it Suzaku} }  
\def\SW{{\it Swift} }
 
\shorttitle{\SU Observations of 7 Unassociated \F Sources} 
\shortauthors{Takahashi et al.}   
 
\title{\SU X-ray Follow-up Observations of Seven Unassociated \F Gamma-ray Sources at High Galactic Latitudes} 
 
\author{Y. Takahashi\altaffilmark{1}, J. Kataoka\altaffilmark{1}, T. Nakamori\altaffilmark{1}, K. Maeda\altaffilmark{1}, R. Makiya\altaffilmark{3}, T. Totani\altaffilmark{3}, C.~C. Cheung\altaffilmark{4}, \L . Stawarz\altaffilmark{2,\,5}, L. Guillemot\altaffilmark{6}, P.~C.~C. Freire\altaffilmark{6}, and I. Cognard\altaffilmark{7}}

\email{s072803523@akane.waseda.jp}

\altaffiltext{1}{Research Institute for Science and Engineering, Waseda University, 3-4-1 Okubo, Shinjuku, Tokyo, 169-8555 Japan} 
\altaffiltext{2}{Department of High Energy Astrophysics, Institute of Space and Astronautical Science (ISAS), Japan Aerospace Exploration Agency (JAXA), 3-1-1 Yoshinodai, Sagamihara, 229-8510 Japan}
\altaffiltext{3}{Department of Astronomy, School of Science, Kyoto University, Sakyo-ku, Kyoto 606-8502}
\altaffiltext{4}{National Research Council Research Associate, resident at Naval Research Laboratory, Washington, DC 20375, USA}
\altaffiltext{5}{Astronomical Observatory, Jagiellonian University, ul. Orla 171, Krak\'ow 30-244, Poland}
\altaffiltext{6}{Max-Planck-Institut f\"ur Radioastronomie, Auf dem H\"ugel 69, 53121 Bonn, Germany}
\altaffiltext{7}{Laboratoire de Physique et Chimie de l'Environnement, LPCE UMR 6115 CNRS, 45071 Orl\'eans Cedex 02, and Station de Radioastronomie de Nan\c{c}ay, Observatoire de Paris, CNRS/INSU, 18330 Nan\c{c}ay, France}
 
\begin{abstract} 
We report on our second-year campaign of X-ray follow-up observations of unidentified \F $\gamma$-ray sources at high Galactic latitudes ($|b|>10^{\circ}$) using the X-ray Imaging Spectrometer onboard the \SU X-ray Observatory. In this second year of the project, seven new targets were selected from the First \F Catalog, and studied with $20-40$\,ks effective \SU exposures. We detected an X-ray point source coincident with the position of the recently discovered millisecond pulsar PSR~J2302+4442 within the $95\%$ confidence error circle of 1FGL~J2302.8+4443. The X-ray spectrum of the detected counterpart was well fit by a blackbody model with temperature of $kT \simeq 0.3$\,keV, consistent with an origin of the observed X-ray photons from the surface of a rotating magnetized neutron star. For four other targets which were also recently identified with a normal pulsar (1FGL~J0106.7+4853) and millisecond pulsars (1FGL~J1312.6+0048, J1902.0$-$5110, and J2043.2+1709), only upper limits in the $0.5-10$\,keV band were obtained at the flux levels of $\simeq 10^{-14}$\,erg\,cm$^{-2}$\,s$^{-1}$. A weak X-ray source was found in the field of 1FGL~J1739.4+8717, but its association with the variable $\gamma$-ray emitter could not be confirmed with the available \SU data alone. For the remaining \F object 1FGL~J1743.8$-$7620 no X-ray source was detected within the LAT $95 \%$ error ellipse. We briefly discuss the general properties of the observed high Galactic-latitude \F objects by comparing their multiwavelength properties with those of known blazars and millisecond pulsars.
\end{abstract} 
 
 
\begin{document} 
\section{Introduction} 
 
Since its successful launch in 2008 June, the Large Area Telescope (LAT) onboard the {\it Fermi} Gamma-ray Space Telescope \citep{atw09} has enabled many important breakthroughs in the understanding of the origin of high energy $\gamma$-ray emissions of various classes of astrophysical objects. The number of detected $\gamma$-ray sources increased dramatically, from 271 objects listed in the 3rd EGRET Catalog \citep[3EG;][]{har99}\footnote{See also \citet{cas08} for the revised catalog of EGRET $\gamma$-ray sources.} to 1873 in the Second \F Catalog \citep[2FGL;][]{2FGL}. About 800 $\gamma$-ray sources included in 2FGL have been identified as blazars \citep{2LAC}, i.e., jetted active galactic nuclei (AGN) characterized by strong relativistic beaming. Other associations included pulsars \citep[e.g.,][]{LATPSR}, high-mass X-ray binaries \citep[e.g.,][]{HMXB}, radio galaxies \citep[e.g.,][]{MAGN}, pulsar wind nebulae \citep[e.g.,][]{pwn}, supernova remnants \citep[e.g.,][]{snr}, globular clusters \citep[e.g.,][]{glbcls}, starburst galaxies \citep[e.g.,][]{sburst}, and distinct objects like the Large Magellanic Cloud \citep{lmc}. However, no obvious counterparts at longer wavelengths have been found for as much as $38\%$ of \F objects so that several hundreds of GeV sources currently remain \emph{unassociated} with any known astrophysical systems. Fortunately, an improved localization error for the \F \citep[typical $95\%$ confidence radii $r_{\rm 95} \sim 0^{\circ}.1-0^{\circ}.2$, and even $0^{\circ}.005-0^{\circ}.01$ for the brightest sources;][]{2FGL}, when compared to that of EGRET (typical $r_{\rm 95} \simeq 0^{\circ}.4-0^{\circ}.7$), allows for much more effective follow-up studies at radio, optical, and X-ray frequencies, which can help to unravel the nature of the unidentified $\gamma$-ray emitters. 
 
In this context, X-ray follow-up observations of unidentified \F objects are of particular importance, since some classes of astrophysical sources of $\gamma$-rays such as AGN are strong X-ray emitters as well, while the others like most of $\gamma$-ray emitting pulsars are faint X-ray sources. Note that assuming the keV--to--GeV emission continuum in the form of a broad-band power-law ($F_{\nu} \propto \nu^{-\alpha_{{\rm x} \gamma}}$), which could be a relatively good zero-order approximation in the case of blazar sources but not necessarily in the case of other classes of $\gamma$-ray emitters,  the monochromatic X-ray flux energy density scales as $[\nu F_{\nu}]_{\rm 1\,keV} = \left({\rm 1\,keV}/{\rm 0.1\,GeV}\right)^{1-\alpha_{{\rm x} \gamma}} \times [\nu F_{\nu}]_{\rm 0.1\,GeV} \simeq 3 \times 10^{-3} \times [\nu F_{\nu}]_{\rm 0.1\,GeV}$ for a relatively flat spectral index of $\alpha_{{\rm x} \gamma} \simeq 0.5$. Hence, if an X-ray counterpart of a bright \F source is characterized by, e.g., $[\nu F_{\nu}]_{\rm 0.1\,GeV} \simeq 10^{-11}$\,erg\,cm$^{-2}$\,s$^{-1}$ and the X-ray--to--$\gamma$-ray power-law emission continuum with the slope $\alpha_{{\rm x} \gamma} \geq 0.5$, such source can be expected to be detectable with modern X-ray instruments such as {\it Chandra}, XMM-{\it Newton}, {\it Swift}, and \SU within reasonable exposure times. In particular, a point source search to the level of $[\nu F_{\nu}]_{\rm 1\,keV} \sim (10^{-14} - 10^{-13})$\,erg\,cm$^{-2}$\,s$^{-1}$ is easily attainable with the X-ray Imaging Spectrometer \citep[XIS;][]{koy07} onboard \SU \citep{mit07} with relatively short exposures of few tens of ksec \citep[e.g.,][]{aka11}. In the case of a positive detection, correlated flux changes at X-ray and $\gamma$-ray frequencies provide an identification. 
The lack of correlated variability, or non-detection of an X-ray counterpart, provide on the other hand only circumstantial evidence regarding the nature of a studied target. Yet in many cases such evidence may be crucial, since the non-detection of an X-ray counterpart despite a long, dedicated observation of a bright \F object may disprove a potential association with given classes of astrophysical sources. That is because, as mentioned above, only a few established high-energy emitters are that bright in $\gamma$-rays but very faint in X-rays \citep[e.g., Geminga pulsar; see the discussion in][]{tho04,mat07}. 

Thus motivated we started a project to investigate the nature of unidentified high Galactic-latitude \F objects through deep X-ray follow-up observations with \SU XIS. The results of the first-year campaign conducted over the span of \SU AO4 were presented in \citet{mae11}. The AO4 program included four steady/weakly variable \F sources from the initial \F Bright Source List \citep[0FGL;][]{0FGL} and can be summarized as follows. The X-ray counterpart for one of the brightest unassociated \F objects, 1FGL~J1231.1$-$1410 (also detected by EGRET as 3EG~J1234$-$1318 and EGR~J1231$-$1412), was found. The X-ray spectrum of the counterpart was well fit by a blackbody model with a temperature of $kT \simeq 0.16$\,keV plus an additional power-law component dominating above 2\,keV photon energies. This power-law component was confirmed in subsequent {\it Swift} and XMM-{\it Newton} exposures. Considering a recent identification of 1FGL~J1231.1$-$1410 with the millisecond pulsar (MSP) PSR~J1231$-$1411 \citep{ran11}, in \citeauthor{mae11} we concluded that the detected thermal X-ray photons originate from the surface of a rotating magnetized neutron star, while the non-thermal X-ray component is most likely produced within the pulsar magnetosphere. In the case of 1FGL~J1311.7$-$3429, two possibly associated X-ray point sources were discovered, one of which is now excluded from the smaller error ellipse of the GeV emitter as catalogued in the 2FGL \citep{2FGL}. The identification of the remaining X-ray counterparts with the respective $\gamma$-ray objects remain uncertain despite a robust determination of the the spectral and variability properties of the X-ray sources. In the case of 1FGL~J1333.2+5056, we found several weak X-ray sources within the \F error circle, and speculated on the AGN nature of the target. Finally, one X-ray point source was detected at the edge of the error ellipse of 1FGL~J2017.3+0603. The physical connection was however viewed as unlikely, since the X-ray source did not coincide with the location of the MSP PSR~J2017+0603 discovered by the Nan\c{c}ay radio telescope which constituted a more highly probable association with the \F object \citep{cog11}. The MSP identification was later indeed confirmed by the detection of the pulsed emission in the \F data, with the same period as the radio pulsations. 

In this paper, we report the results of our second-year campaign, conducted over the span of \SU AO5 (2010 April to 2011 March) which included observations of seven \F sources located at high Galactic latitudes ($|b|>10^{\circ}$). The targets were selected from the First \F Catalog of point sources \citep[1FGL;][]{1FGL} as objects \emph{unidentified} at the time of writing of the \SU AO5 proposal. Since then however, four of the selected targets have been associated with MSPs: 1FGL~J1902.0$-$5110 with PSR~J1902$-$5105 \citep{cam11}, 1FGL~J2043.2+1709 with PSR~J2043+1711 \citep{gui11}, 1FGL~J2302.8+4443 with PSR~J2302+4442 \citep{cog11}, and 1FGL~J1312.6+0048 with PSR~J1312+00 \citep{2FGL}. And moreover, 1FGL~J0106.7+4853 have very recently been associated with a normal pulsar PSR~J0106+4855 \citep{ple11}. The XMM-{\it Newton} and {\it Swift} satellites detected the weak X-ray counterpart of the MSP in 1FGL~J2302.8+4443 \citep{cog11}. In the following, our new \SU observations and data reduction procedure are described in section~\ref{sec:obs}. The analysis results are given in section~\ref{sec:results}, and discussed further in section~\ref{sec:disc_and_conc} in the context of multiwavelength studies of unidentified \F objects. 
 
\section{{\it Suzaku} Observations and Data Analysis} 
\label{sec:obs} 
 
\subsection{Observations and Data Reduction} 
\label{sec:obs-reduct} 
 
We observed seven unidentified \F sources with the \SU XIS. These seven targets were chosen from the 1FGL catalog according to the following selection criteria: (i) no association claimed at the time of the submission of the \SU proposal, (ii) sources located more than $\pm 10^{\circ}$ away from the Galactic plane, and (iii) the detection significance in the LAT $\gamma$-ray band ($>100$\,MeV) exceeding $14 \sigma$ in the 1FGL catalog. The thus selected targets are listed in Table\,\ref{tab:obs_log} together with the corresponding \SU observation logs and the 1FGL $\gamma$-ray fluxes and photon indices.  
 
The observations were conducted with three XIS detector and the Hard X-ray Detector \citep[HXD;][]{kok07,tak07}. The XIS detector is composed of four CCD cameras. One of the four CCD cameras (XIS1) is back-illuminated CCD and the others (XIS0, XIS2, and XIS3) are front-illuminated CCDs. The operation of XIS2 ceased in 2006 November because of the contamination by a leaked charge. Since none of the studied sources have been detected with the HXD, below we describe the analysis of only the XIS data. The XIS was operated in pointing mode and the normal clocking mode, combined with the two editing modes $3 \times 3$ and $5 \times 5$ for five targets, and only one editing mode $3 \times 3$ for 1FGL~J2302.8+4443 and 1FGL~J1312.6+0048 because of the telemetry limit. 
 
We conducted all the data reduction and analysis with \texttt{HEADAS} software version 6.9 and the calibration database (CALDB) released on 2010 July 30. First, we combined the cleaned event data of the two editing modes using \texttt{xselect}. Then we removed the data corresponding to the epoch of low-Earth elevation angles (less than $5^{\circ}$), as well periods (and 60\,sec after) when the \SU satellite was passing through the South Atlantic Anomaly (SAA). Moreover, we also excluded the data obtained when the \SU satellite was passing through the low Cut-Off Rigidity (COR) of below 6\,GV. Finally we removed hot and flickering pixels using \texttt{sisclean} \citep{day98}. 
 
\subsection{Data Analysis} 
\label{sec:danalys} 
 
X-ray images of each target were extracted from the two operating front-illuminated CCDs (XIS0 and XIS3). In the image correction procedure we applied a `non X-ray background' subtraction, an exposure correction, and a vignetting correction \citep[for details see][]{mae11}. We then combined the images from both CCDs and smoothed the thus obtained final maps using a Gaussian function with $\sigma = 0'.28$. The resulting images are presented and discussed in the next section\,\ref{sec:results}. Although all the $\gamma$-ray targets were initially selected from the 1FGL catalog, in all the corresponding figures, thick green ellipses denote the more precise $95\%$ position errors from the 2FGL catalog \citep{2FGL} as described in section\,\ref{sec:obs-reduct}. 

For the further analysis, we selected source regions around each detected X-ray source within the 2FGL error ellipses. The radii of the source extraction regions, denoted in the figures below by thin green circles, were set as $1'$, unless otherwise stated. The corresponding background regions with radii of $3'$ were taken from the low count rate area in the same XIS chips (dashed green circles). We set the detection threshold for X-ray sources at $4\sigma$, based on the signal-to-noise ratio defined as a ratio of the excess events above a background to its standard deviation assuming a Poisson distribution. The X-ray source positions and the corresponding errors were estimated by 2D Gaussian fits, as summarized in Table\,\ref{tab:results}. 
 
For the timing analysis, light curves from the front-illuminated (XIS0, XIS3) and back-illuminated (XIS1) CCDs were combined; the corresponding backgrounds were subtracted using \texttt{lcmath}. The light curves constructed in this way provide the net-count rates. To quantify possible flux variations, the $\chi^2$ test was applied to each light curve using \texttt{lcstats}. For the X-ray spectral analysis, we generated the RMF files for the detector response and the ARF files for the effective area using \texttt{xisrmfgen} and \texttt{xissimarfgen} \citep{ish07}. In order to improve the statistics, we added X-ray counts from the two front-illuminated CCDs, using \texttt{mathpha} with no error propagation so that the resulting data follow a Poisson distribution, and then combined the response files using the \texttt{marfrmf} and \texttt{addrmf} commands. In the case of the $\gamma$-ray targets with no detected X-ray counterparts within the 2FGL error ellipses, we calculated $90\%$ confidence level flux upper limits at the positions of the $\gamma$-ray emitters assuming an absorbed power-law model. Uncertainties of the model spectral parameters are computed at the $90\%$ confidence levels. The results of the spectral fitting are summarized in Table\,\ref{tab:fits}, and discussed below in more detail. 

\section{Results} 
\label{sec:results} 

In this section, we first present the results of the new \SU observations in the order of R.A. for the sources for which we detected X-ray counterparts, followed by the analysis results of the remaining targets with no detected X-ray counterparts. All seven targets have also recently been observed by {\it Swift}, and below we briefly compare the results of the \SU and {\it Swift} observations. We search for radio, infrared and optical counterparts for all the detected X-ray sources using the NRAO VLA Sky Survey (NVSS) catalog \citep{con98}, the Two Micron All Sky Survey (2MASS) point source catalog \citep{2MASS}, and the USNO-B1.0 catalog \citep{usb1}.

\subsection{1FGL~J1312.6+0048} 
\label{sec:J1312} 

We discovered one X-ray point source inside the 2FGL error ellipse of 1FGL~J1312.6+0048, at [RA, Dec]=[198$^{\circ}$.235(2), 0$^{\circ}$.835(2)], designated hereafter as Suzaku J1312+0050, with a detection significance of $9\sigma$ (total of 178 net source counts from three detectors). This X-ray source has recently been detected also by \SW X-ray Telescope \citep[XRT;][]{xrt} at [RA, Dec] = [265.4229, 0.8348] with 90\% position error radius r$_{90\%}$ = 6$''$.8. Since the position accuracy of \SW XRT\footnote{See \SW Technical Handbook (\texttt{http://heasarc.nasa.gov/docs/swift/proposals/appendix\_f.html})} ($\sim$ 5$''$) is better than that of \SU XIS \citep[$\sim$ 19$''$;][]{uch08}, we searched for optical and radio counterparts of the X-ray source using the \SW position. We found one possibly related optical emitter USNOB 0908-0218088 at about 5$''$away from the X-ray source ([RA, Dec] = [198.241972(7), 0.83448(5)], and magnitudes B2 = 19.80, R2 = 19.32, and I = 18.11), while no radio and infrared counterparts were discovered.

The light curve of Suzaku J1312+0050 with a time bin of 5774\,s and its spectrum are presented in Figures\,\ref{fig:J1312_X_lc} and \ref{fig:J1312_X_spec}, respectively. The X-ray spectrum when fitted with an absorbed power-law model returned negligible hydrogen column density. We therefore fixed $N_{\rm H} = 0$ and repeated a power-law fit obtaining the photon index of $\Gamma = 1.9^{+0.4}_{-0.3}$ with $\chi^{2}$/d.o.f.=9.9/13. The derived energy flux in the $2-10$\,keV photon energy range is $8.0^{+2.9}_{-2.6} \times 10^{-14}$\,erg\,cm$^{-2}$\,s$^{-1}$. To test for flux variability, we performed a $\chi^{2}$ fit with a constant count rate of $1.02 \times 10^{-1}$\,ct\,s$^{-1}$ resulting in $\chi^{2}$/d.o.f. = 14.5/7. This indicates that the X-ray source is variable on the timescale of a few hours, with a probability of $\simeq 96\%$. 

The targeted $\gamma$-ray object has recently been associated with the MSP PSR~J1312+00 \citep{2FGL}. Unfortunately, the position of the radio pulsar is still not available publicly, and hence at this moment we cannot claim nor reject the coincidence of PSR~J1312+00 with the detected X-ray source. The variability revealed by the \SU data, along with the presence of an optical counterpart, implies however that the detected X-ray source is not likely associated with the MSP PSR~J1312+00. For this reason we have evaluated the $90 \%$ confidence upper limit to the $0.5-2$\,keV and $2-10$\,keV fluxes at a random position within the 2FGL error ellipse of 1FGL~J1312.6+0048 (excluding Suzaku J1312+0050), which should correspond (at least roughly) to the upper limits for the X-ray emission of PSR~J1312+00. Assuming an absorbed power-law model with $n_{\rm H}$ = 2.1 \citep[this value is taken from the LAB Survey of Galactic HI;][]{kal05} and photon index $\Gamma = 2$, these read as $< 1.3 \times 10^{-13}$\,erg\,cm$^{-2}$\,s$^{-1}$ ($0.5-2$\,keV) and $< 9.3 \times 10^{-14}$\,erg\,cm$^{-2}$\,s$^{-1}$ ($2-10$\,keV), respectively. The implied ratio of the $0.1-100$\,GeV and $2-10$\,keV fluxes $F_{\rm \gamma}/F_{\rm X}$ $>$ 211 would be then in agreement with the pulsar association of 1FGL~J1312.6+0048 \citep[see][]{mar11}.

\subsection{1FGL~J1739.4+8717}
\label{sec:J1739}

We detected one X-ray source within the 2FGL error ellipse of 1FGL~J1739.4+8717 (designated as Suzaku J1742+8715). The detection significance is 7$\sigma$ (total of 204 net source counts from three detectors). The XIS image of Suzaku J1742+8715 seems to be relatively diffuse. However, \SW XRT has detected recently the same X-ray emitter as a point source at [RA, Dec] = [265.430, 87.245] with 90\% position error radius r$_{90\%}$ = 6$''$.9. Hence we conclude that in a relatively short \SU exposure (16.7 ksec) the apparently diffuse structure of the object is just an artifact of a low photon statistics. We found several weak optical sources coinciding positionally with the X-ray source, but no radio counterpart.

We extracted the source photons from a $2'$ radius circle around Suzaku J1742+8715 and the background photons from a $3'$ radius circle. The resulting X-ray spectrum and X-ray light curve with the time bin of 5760\,s are shown in Figures\,\ref{fig:J1739_xspec} and \ref{fig:J1739_lc}, respectively. A constant fit to the light curve of the X-ray source returned $\chi^{2}$/d.o.f. = 1.9/6, indicating that the flux was steady during the \SU exposure with the $\chi^{2}$ probability of $> 93 \%$. The spectrum was initially fitted by an absorbed power-law model ($\chi^{2}$/d.o.f. = 11.4/12) and this model returned $N_{\rm H}$ = 0. We then refit the spectrum after fixing the $N_{\rm H}$ value to zero and obtained a photon index of $\Gamma = 2.1\pm^{+0.5}_{-0.4}$ and the unabsorbed $2-10$\,keV flux $3.6^{+2.0}_{-1.5}$\,erg\,cm$^{-2}$\,s$^{-1}$.

We found one relatively bright radio source inside the 2FGL error region of 1FGL~J1739.4+8717, namely NVSS~J173722+871744 ($\sim 60$\,mJy at 1.4\,GHz). Its position is marked with a green rhombus in Figure~\ref{fig:J1739_Ximage}. This radio source is detected at several frequencies \citep[151\,MHz and 325\,MHz; see][respectively]{bal85,ren97}, but has no obvious X-ray counterpart in our \SU data. Assuming a power-law form of the radio continuum ($F_{\nu} \propto \nu^{-\alpha_{\rm r}}$), we calculate the spectral index of NVSS\,J173722+871744 as $\alpha_{\rm r} = 0.2$ $\pm$ 0.1 from the archival data. This is, in fact, a typical radio spectral index for blazar sources \citep[e.g.,][]{sow05}. Moreover, we also found an optical and infrared counterpart, USNOB 1772-0020476 ([RA, Dec] = [264.35499(2), 87.29532(2)]), which is located only 2$''$.2 away from the NVSS source with optical magnitudes, B2 = 19.30, R2 = 17.61, and I = 17.07, and 2MASS J17372480+8717433 ([RA, Dec] = [264.3533(1), 87.2953(1)], and magnitudes J = 16.2, H = 15.5, and K = 15.8). This allowed us to calculate radio-to-optical and optical-to-X-ray spectral indices as $\alpha_{\rm ro} = 0.40$ $\pm$ 0.01 and $\alpha_{\rm ox} > 1.77$, with the latter utilizing the X-ray flux upper limit (see Figure~\ref{fig:J1739sed}). These indices are consistent with those of so-called intermediate synchrotron peaked blazars \citep[see Fig. 7 in][]{2LAC}.

\subsection{1FGL~J2043.2+1709} 
\label{sec:J2043} 

We found one X-ray point source within the 2FGL error ellipse of 1FGL~J2043.2+1709 at the respective position [RA, Dec]=[310$^{\circ}$.801(2), 17$^{\circ}$.171(2)], designated as Suzaku J2043+1710, and a second source located on the edge of the ellipse at [RA, Dec]=[310$^{\circ}$.822(3), 17$^{\circ}$.133(3)], designated as Suzaku J2043+1707 (hereafter srcA and srcB for short, respectively). These two X-ray objects are not recorded in any available X-ray source catalogs. \SW XRT has also detected the same two X-ray emitters. Moreover, \SW Ultraviolet/Optical Telescope \citep[UVOT;][]{uvot} discovered a possible UV counterpart of srcA at [RA, Dec] = [310.80314(6), 17.17286(7)], for which we found an association with the optical source USNOB 1071-0645302 ([RA, Dec] = [310.80315(7), 17.17300(4)], and magnitudes B2 = 19.89, R2 = 20.6). No radio and infrared counterparts for srcA or srcB were found. The X-ray image of the targeted field is shown in Figure\,\ref{fig:J2043_image}. The radio position of the MSP PSR~J2043+1711 recently associated with the 1FGL~J2043.2+1709 is marked in the figure with a green cross centered at [RA,Dec]=[310$^{\circ}$.8370129(2), 17$^{\circ}$.1913744(3)].
 
To extract the spectra and light curves of the detected X-ray sources, we set the source regions and the background region as indicated in Figure\,\ref{fig:J2043_image}. The detection significances of both sources were calculated to be $16 \sigma$ (total of 397 net source counts from three detectors) and $8 \sigma$ (total of 155 net source counts from three detectors), respectively. No X-ray counterpart of PSR~J2043+1711 was detected in our \SU exposure. The light curves of srcA and srcB with time bins of 5760\,s are given in Figure\,\ref{fig:J2043_X_lc}, and the corresponding spectra are presented in Figure\,\ref{fig:J2043_spectrum}. Both light curves can be well fit by constant count rates ($\chi^{2}$/d.o.f. = 8.5/9 and $\chi^{2}$/d.o.f. = 2.6/9 for srcA and srcB, respectively). 

An absorbed power-law model provided the best fit to the spectrum of srcA, returning the photon index $\Gamma = 1.67^{+0.16}_{-0.15}$ and $\chi^{2}$/d.o.f = 20.4$/$19. In the fit, the value of $N_{\rm H}$ was frozen at the Galactic value in the direction of the target taken from the LAB Survey of Galactic HI, i.e., $6.63 \times 10^{20}$\,cm$^{-2}$ \citep{kal05}. We note that the fit with the absorption column density set free returned the value for $N_{\rm H}$ consistent with the Galactic one. The derived unabsorbed X-ray flux of srcA in the $2-10$\,keV photon energy range is $2.1^{+0.4}_{-0.4} \times 10^{-13}$\,erg\,cm$^{-2}$\,s$^{-1}$. Similarly, the X-ray spectrum of srcB was initially fitted with an absorbed power-law model. However, this model fit returned negligible value of $N_{H}$ and therefore we fixed $N_{\rm H}$ = 0. The best fit model with $\chi^{2}$/d.o.f. = 8.6/10 returned then the photon index $\Gamma=1.6^{+0.4}_{-0.3}$ and the unabsorbed $2-10$\,keV flux $9.1^{+3.5}_{-3.0} \times 10^{-14}$\,erg\,cm$^{-2}$\,s$^{-1}$.

Assuming a power-law model with photon index $\Gamma = 2$, we calculated upper limits ($90\%$ confidence) in the $0.5-2$\,keV and $2-10$\,keV bands at the position of PSR~J2043+1711 as $< 2.4 \times 10^{-14}$\,erg\,cm$^{-2}$\,s$^{-1}$ and $< 3.6 \times 10^{-14}$\,erg\,cm$^{-2}$\,s$^{-1}$, respectively. This gives the ratio of the $0.1-100$\,GeV and $2-10$\,keV fluxes $F_{\rm \gamma}/F_{\rm X}$ $>$ 864 which is consistent with the values claimed for the GeV-detected MSP \citep{mar11}.

\subsection{1FGL~J2302.8+4443} 
\label{sec:J2302} 

One X-ray point source was discovered inside the 2FGL error ellipse of 1FGL~J2302.8+4443 at the position of [RA, Dec]=[345$^{\circ}$.695(3), 44$^{\circ}$.707(2)] (designated as Suzaku J2302+4442). The detection significance of this X-ray source was calculated as $5.35 \sigma$ (total of 121 net source counts from three detectors). As shown in Figure\,\ref{fig:J2302_image}, the detected X-ray source coincides with the radio position of the MSP PSR~J2302+4442 (green cross centered at [RA, Dec]=[345$^{\circ}$.695748(3), 44$^{\circ}$.706136(1)]). This MSP was recently claimed to be associated with 1FGL~J2302.8+4443. No optical and infrared counterparts of the pulsar were found. In the analysis of the \SU data, the source and the background regions were set as indicated in the Figure. The light curve of the X-ray counterpart with the time bin of 11520\,s is presented in Figure\,\ref{fig:J2302_X_lc}. The applied $\chi^{2}$ test assuming a constant count rate gave $\chi^{2}$/d.o.f= 4.98/5, indicating that the X-ray flux was steady during the \SU exposure. 
 
The XIS spectrum of the detected X-ray source is shown in Figure\,\ref{fig:J2302_Xspec}. Initially, we fitted the spectrum with an absorbed power-law model. This model returned an extremely soft continuum characterized by a photon index of $\Gamma = 4.5^{+2.1}_{-1.7}$, and a rather high value of $N_{\rm H} = 0.56^{+0.68}_{-0.34} \times 10^{22}$\,cm$^{-2}$. This absorption value is in excess of the Galactic column density in the direction of the target, namely $0.132 \times 10^{22}$\,cm$^{-2}$, as determined by \citet{kal05} from the LAB survey of Galactic HI. Next we applied an absorbed black body model but the model fit returned $N_{\rm H}$ consistent with zero. We therefore fixed $N_{\rm H} =0$ and obtained the best-fit ($\chi^{2}$/d.o.f. = 3.0/11) temperature of $kT \simeq 0.31^{+0.06}_{-0.05}$\,keV (see Figure\,\ref{fig:J2302_Xspec}). The observed X-ray flux of the source is $2.6^{+0.6}_{-0.5} \times 10^{-14}$\,erg\,cm$^{-2}$\,s$^{-1}$ in the $0.5-2$\,keV range and 3.3 $^{+2.7}_{-2.0}$ $\times 10^{-15}$\,erg\,cm$^{-2}$\,s$^{-1}$ in the $2-10$\,keV range. The ratio between $\gamma$-ray and $2-10$\,keV fluxes for 1FGL~J2302.8+4443/Suzaku J2302+4442 reads as $F_{\rm \gamma}/F_{\rm X}$ $=$ 14531. This ratio is relatively high but still consistent with those of \F/LAT pulsars reported in \citet{mar11}.

\subsection{Other Sources}
\label{sec:others}

No X-ray sources were detected in our \SU exposures inside the 2FGL error ellipses of the remaining $\gamma$-ray objects: 1FGL~J0106.7+4853, 1FGL~J1743.8$-$7620, and 1FGL~J1902.0$-$5110. The corresponding X-ray images of the targets are shown in Figures\,\ref{fig:J0106_X_image}, \ref{fig:J1743_X_image}, and \ref{fig:J1902_X_image}. Out of these three sources, 1FGL~J1902.0$-$5110 has been recently associated with the MSP PSR~J1902$-$5105. Even more recently, after the submission of this paper, also 1FGL~J0106.7+4853 has been identified with a new $\gamma$-ray pulsar PSR~J0106+4855 \citep{ple11}. Thus only 1FGL~J1743.8$-$7620 remains at the moment unassociated. We calculated an X-ray upper limit in the $2-10$\,keV band for the latter source assuming X-ray emission from a point source located at the center of 2FGL error region, obtaining in this way the energy flux $< 4.5 \times 10^{-14}$\,erg\,cm$^{-2}$\,s$^{-1}$. In the case of 1FGL~J1902.0$-$5110, we set the X-ray extraction region as a $1'$ radius circle around the radio position of PSR~J1902$-$5105. The resulting X-ray upper limits are calculated as $< 2.0 \times 10^{-14}$\,erg\,cm$^{-2}$\,s$^{-1}$ ($0.5-2$\,keV) and $< 2.5 \times 10^{-14}$\,erg\,cm$^{-2}$\,s$^{-1}$ ($2-10$\,keV) by assuming an absorbed power-law model with photon index $\Gamma = 2$. Finally, X-ray upper limits for 1FGL~J0106.7+4853 were calculated for a circular extraction region with radius 1' around the position of PSR~J0106+4855, obtaining $<7.7$ $\times$ 10$^{-15}$\,erg\,cm$^{-2}$\,s$^{-1}$ ($0.5-2$\,keV) and $<1.0$ $\times$ 10$^{-14}$\,erg\,cm$^{-2}$\,s$^{-1}$ ($2-10$\,keV). For these two pulsars, the ratio of the $0.1-100$\,GeV and $2-10$\,keV fluxes are $F_{\rm \gamma}/F_{\rm X}$ $>$ 976 and $F_{\rm \gamma}/F_{\rm X}$ $>$ 2640, respectively. These flux ratios are consistent with the ratios of \F pulsars reported in \citet{mar11}. The results of the \SW observations of all of the three targeted $\gamma$-ray emitters are consistent with the \SU results.

\section{Discussion and Conclusions}
\label{sec:disc_and_conc}

In this paper, we report on the results of X-ray follow-up observations of seven bright \F sources at high Galactic latitudes ($|b|>10^{\circ}$) using \SU XIS. We discovered the X-ray counterpart of 1FGL~J2302.8+4443 coinciding with the position of the MSP PSR~J2302+4442 recently claimed to be associated with the $\gamma$-ray emitter. We did not however, detect X-ray counterparts for the other four \F objects similarly identified with a normal pulsar (1FGL~J0106.7+4853) and MSPs, namely for 1FGL~J1312.6+0048, 1FGL~J1902.0$-$5110 and 1FGL~J2043.2+1709. (In a few cases the X-rays sources have been detected within the 2FGL error ellipses, but none at the positions of the pulsars.) A relatively weak X-ray source was found inside the 2FGL error region of 1FGL~J1739.4+8717. Finally, no candidate for the X-ray counterpart was detected for the remaining object 1FGL~J1743.8$-$7620.

Including our previous observations of 1FGL~J1231.1$-$1410, 1FGL~J1311.7$-$3429, 1FGL~J1333.2+5056, and 1FGL~J2017.3+0603 reported in \citet{mae11}, our sample of high Galactic-latitude \F objects initially selected as unidentified and studied with \SU consists now of eleven targets. For eight of these, we have detected single or multiple X-ray sources within the LAT error ellipses. Over the time period when the \SU observations were being obtained, six targets from the $\gamma$-ray sample were found to be associated with MSPs, one target (1FGL~J0106.7+4853) has been associated with a normal pulsar, and one source (1FGL~J1333.2+5056) has been classified as an AGN candidate, all in agreement with the gathered X-ray data. Still, four objects from the list remain unidentified. As argued below, one of these four, 1FGL~J1739.4+8717, is quite likely a high-redshift blazar. 

The source 1FGL~J1739.4+8717 was characterized by an enhanced flux level within the LAT photon energy range during the the first seven months of \F operation\footnote{\texttt{http://heasarc.gsfc.nasa.gov/FTP/fermi/data/lat/catalogs/source/lightcurves/2FGLJ1738.9+8716.png}}. After that time, the activity of the $\gamma$-ray emitter decreased. The photon index in the LAT energy band is $\Gamma_{\gamma}$ = 2.1 $\pm$ 0.1 \citep[where dN/dE $\propto$ E$^{-\Gamma_{\gamma}}$ is the differential photon flux; ][]{2FGL}, which is a typical value for the $\gamma$-ray spectra of BL Lac type blazars \citep[see][]{2LAC}. Importantly, as written in section~\ref{sec:J1739} one relatively bright radio source, NVSS\,J173722+871744, is located inside the 2FGL error region of 1FGL~J1739.4+8717 (see Figure\,\ref{fig:J1739_Ximage}). With a typical radio spectral index for blazar sources and radio-to-optical and optical-to-X-ray spectral indices that are consistent with blazar broadband spectrum, it is quite likely that 1FGL~J1739.4+8717 is indeed associated with a distant blazar currently characterized by an activity level low enough so that its X-ray emission was below the detection limit of the XIS instrument ($\sim 10^{-15}$\,erg\,cm$^{-2}$\,s$^{-1}$) at the time of the performed \SU observations.

In general, unidentified sources constituted a large fraction of the population of $\gamma$-ray emitters detected by EGRET  ($\sim 60\%$ in 3EG), and at present about 31 $\%$ of \F sources in 2FGL catalog remain unassociated (specifically 273 sources at high Galactic latitudes ($| b | > 10^{\circ}$) and 303 sources at low Galactic latitudes $| b | < 10^{\circ}$). Those located at the lowest Galactic latitudes ($|b| < 5^{\circ}$) are most widely expected to be associated with local systems such as molecular clouds, supernova remnants, massive stars, high-mass X-ray binaries, radio quiet pulsars, and pulsar wind nebulae \citep[e.g.,][]{kaa96,yad97,rom99}. In particular, half a dozen of the brightest 3EG sources in the Galactic plane were identified as young pulsars \citep{tho99}, despite the relatively poor localization of the EGRET sources and the source confusion complicated substantially the identification procedure. On the other hand, most of the unassociated 3EG sources at high Galactic latitudes ($|b| > 10^{\circ}$) were later identified as blazars \citep{sow03,sow04}. Pulsars were therefore expected to be found mainly among GeV emitters at low Galactic latitudes, while blazars were supposed to constitute the main population of GeV emitters at high Galactic latitudes. But the $\gamma$-ray--bright pulsars were also expected to be found at intermediate Galactic latitudes \citep[$5^{\circ} < |b| < 73^{\circ}$;][]{cra06}. The identification of a number of \F objects located above the Galactic plane ($| b | > 10^{\circ}$) with such systems (predominantly with MSPs) confirmed these expectations \citep[see, e.g.,][and the discussion in \citealt{mae11}]{ran10}. 

In Figure\,\ref{fig:ratio} we plot the X-ray--to--$\gamma$-ray energy flux density ratios ($F_{2-10\,{\rm keV}}/F_{0.1-100\,{\rm GeV}}$) versus radio--to--$\gamma$-ray energy flux density ratios ($F_{1.4\,{\rm GHz}}/F_{0.1-100\,{\rm GeV}}$) for the \F objects from our \SU sample discussed here (blue circles) with the radio data available in the literature \citep[see Table\,\ref{tab:fits}]{1FGL,con98}. These can be compared with the analogous ratios evaluated for bright \F objects identified with blazars and MSPs (denoted in the figure by red crosses and squares, respectively). We remind the reader that the blazar class includes flat spectrum radio quasars (FSRQs) and BL Lacertae objects (BL Lacs). In addition, in the figure we plot the two targets discussed in \citet{mae11}, namely 1FGL~J1333.2+5056 most likely associated with an AGN, and a peculiar object 1FGL~J1311.7$-$3429 (pink stars). As shown, the blazar and MSP populations are clearly separated in the constructed flux ratio plane. Also, four objects from our sample which have recently been associated with a normal pulsar (1FGL~J0106.7+4853) and MSPs (1FGL~J1312.6+0048, 1FGL~J2043.2+1709, and 1FGL~J2302.8+4443) occupy the same region in the analyzed parameter space as the previously known MSPs detected in the GeV range\footnote{In the case of 1FGL~J1312.6+0048, the X-ray--to--$\gamma$-ray energy flux density ratio shown in Figure\,\ref{fig:ratio} is evaluated assuming the association of the $\gamma$-ray sources with the MSP PSR~J1312+00. That is, the \SU XIS upper limit derived at the position of the pulsar is considered, and not the X-ray flux of the \SU source detected within 2FGL error region.}. On the other hand, in the case of 1FGL~J1739.4+8717 the evaluated energy flux density ratios --- which are very similar to those characterizing 1FGL~J1333.2+5056 --- are consistent with the blazar identification proposed above, if only NVSS\,J173722+871744 is considered as the true counterpart of the $\gamma$-ray emitter. 

In all, we conclude that the gathered \SU XIS data together with the broad-band properties of the analyzed \F objects are in agreement with the identification of most of them as MSPs. Yet a few cases in the analyzed sample (1FGL~J1739.4+8717, 1FGL~J1333.2+5056) constitute quite probable associations with AGN (high-redshift blazars). Finally, the nature of the remaining few targets (like 1FGL~J1311.7$-$3429) is still an open question, although, as inferred from Figure \ref{fig:ratio}, the MSP identification seems more viable than the blazar one. In the near future, we are further continuing our X-ray studies during the \SU AO6 cycle, focusing on both new targets (1FGL~J0103.1+4840, 1FGL~J1946.7$-$5404, and 1FGL~J2339.7$-$0531), but also performing ultra-deep exposures on particularly intriguing sources like 1FGL~J1311.7-3429.

Finally, let us comment in more detail on the case of 1FGL~J2302.8+4443. This object, as already mentioned above, has recently been associated with the millisecond pulsar PSR~J2302+4442 discovered by the Nan\c{c}ay radio telescope \citep{cog11}. The rotation period of the pulsar is $P \simeq 5.19$\,ms, the spin-down luminosity is $\dot{E} \simeq 3.74 \times 10^{33}$\,erg\,s$^{-1}$, and the characteristic age can be evaluated as $\tau \simeq 6.2$\,Gyr. \citet{cog11} reported also on the detection of the X-ray counterpart of the pulsar with XMM-{\it Newton}, with the unabsorbed $0.5-3$\,keV flux of $3.1^{+0.4}_{-0.4} \times 10^{-14}$\,erg\,cm$^{-2}$\,s$^{-1}$. This is consistent with our \SU detection (the re-calculated flux in the same photon energy range $2.9^{+1.1}_{-1.2} \times 10^{-14}$\,erg\,cm$^{-2}$\,s$^{-1}$). Anticipating the distance of the pulsar $d \simeq 1.18$\,kpc which was inferred from the NE2001 model of Galactic free electron density \citep{cor01}, the X-ray luminosity of PSR~J2302+4442 can therefore be evaluated roughly as $L_{\rm x} \sim 3 \times 10^{30}$\,erg\,s$^{-1}$. This, together with $\dot{E}$ provided above, is then in good agreement with the scaling relation between the X-ray and spin-down luminosities $L_{\rm x} \sim 10^{-3} \times \dot{E}$ established for MSPs \citep{bec97,gae06,zha07}, although the inferred distance of PSR~J2302+4442 is indicated to be smaller by a factor of four considering an unphysically high $\gamma$-ray efficiency and, instead, assuming average efficiency of $\gamma$-ray MSPs $\sim$ 10 \% \citep{cog11}.

The spectrum of the X-ray counterpart of PSR~J2302+4442/1FGL~J2302.8+4443 was well fit by a blackbody model, and this is again in agreement with the idea that the observed X-ray photons originate from thermal emission from the surface of a rotating magnetized neutron star. Because of the limited photon statistics in the higher energy range of XIS, we could not however confirm the presence of a non-thermal component above 2\,keV \citep[clearly detected in the case of 1FGL~J1231.1$-$1410 by][]{mae11}. Interestingly, for the isolated pulsars as old as PSR~J2302+4442 (characteristic age of about 6\,Gyr), the surface temperature of the neutron star is expected to be $T \lesssim 10^{5}$\,K \citep{nam87,pag92}. Our detection of the X-ray counterpart indicated $kT \simeq 0.31 \pm 0.03$\,keV, i.e. $T \simeq 3.6 \times 10^{6}$\,K instead. Some reheating process is therefore required, possibly related to the impact of relativistic particles on polar caps \citep[and references therein]{bec97}.

C.C. Cheung's work at NRL is sponsored by NASA DPR S-15633-Y. \L .S. is grateful for the support from Polish MNiSW through the grant N-N203-380336. We thank the anonymous referee for a careful reading of the manuscript and useful suggestion which helped to improve the paper.

\begin{table}[m] 
\footnotesize 
\caption{\SU Observation Logs and $\gamma$-ray properties of the targets} 
\begin{center} 
\begin{tabular}{ccccccccc} 
\hline\hline 
Name & OBS ID & \multicolumn{2}{c}{Pointing Direction$^{a}$} & Observation & 
 Effective &  $\gamma$-ray photon & $\gamma$-ray flux$^{b}$ \\  
 & & & & start & exposure & index$^{b}$ & $F_{0.1-100 {\rm GeV}}$ \\ 
     &        & \multicolumn{1}{c}{RA [deg]}& \multicolumn{1}{c}{DEC [deg]} & (UT) & [ksec] &  & [erg cm$^{-2}$ s$^{-1}$] \\ 
\hline
1FGL J0106.7$+$4853 & 705010010 & 16.6433 & 48.9425 & 2010/07/15 09:53:44 & 19.7 & 2.05 & 2.64 $\times$ 10$^{-11}$ \\ 
1FGL J1312.6$+$0048 & 705011010& 198.1860 & 0.8370 & 2011/01/17 04:38:43 & 17.5 & 1.99 & 1.96 $\times$ 10$^{-11}$ \\ 
1FGL J1739.4$+$8717 & 705012010 & 264.8730 & 87.2900 & 2010/04/26 23:41:56 & 16.7 & 2.12 & 2.92 $\times$ 10$^{-11}$\\ 
1FGL J1743.8$-$7620 & 705013010 & 265.9610 & $-$76.3420 & 2010/04/14 00:16:05 & 34.0 & 2.14 & 2.83 $\times$ 10$^{-11}$ \\ 
1FGL J1902.0$-$5110 & 705014010 & 285.5220 & $-$51.1700 & 2010/04/13 06:37:43 & 38.5 & 2.10 & 2.46 $\times$ 10$^{-11}$ \\ 
1FGL J2043.2$+$1709 & 705015010 & 310.8220 & 17.1640 & 2010/05/03 19:23:03 & 17.7 & 2.13 & 3.15 $\times$ 10$^{-11}$\\ 
1FGL J2302.8$+$4443 & 705016010 & 345.7070 & 44.7230 & 2010/06/26 05:52:36 & 35.6 & 2.04 & 4.81 $\times$ 10$^{-11}$\\ 
\hline
\end{tabular}
\end{center}
$^{a}$ The pointing directions are the values of planned target coordinates.\\
$^{b}$ These values are taken from 1FGL catalog \citep{1FGL}.
\label{tab:obs_log}
\end{table}

\begin{table}[m]
\footnotesize 
\caption{Results of the \SU observations of the selected \F sources. Positions of detected X-ray sources are presented.}
\begin{center} 
\begin{tabular}{cccc} 
\hline\hline 
\multicolumn{2}{c}{Name} & \multicolumn{2}{c}{Position} \\  
 & Suzaku detection & RA[deg] & DEC[deg]  \\ 
\hline 
1FGL J0106.7+4853 & - & - & -\\ 
1FGL J1312.6+0048 & SUZAKU~J1312+0050 & 198.235(2) & 0.835(2)\\ 
1FGL J1739.4+8717 & source$^{a}$ & - & - \\ 
1FGL J1743.8$-$7620 & - & - & - \\
1FGL J1902.0$-$5110 & - & - & - \\
1FGL J2043.2+1709 & SUZAKU~J2043+1710 & 310.801(2) & 17.171(2) \\ 
 & SUZAKU~J2043+1707 & 310.822(3) & 17.133(3)\\ 
1FGL J2302.8+4443 & PSR~J2302+4442 & 345.695(3) & 44.707(2)\\ 
\hline 
\end{tabular} 
\end{center} 
$^{a}$We could not determine the exact position of the source because of extended source image (see section 3.2).
\label{tab:results} 
\end{table}

\begin{table}[m] 
\footnotesize 
\caption{Results of the analysis of the \SU/XIS data and radio fluxes of the sources} 
\begin{center} 
\begin{tabular}{cccccccc} 
\hline\hline 
\multicolumn{2}{c}{Name} & $N_{\rm H}$ & Model & Model & X-ray flux  & Radio flux\\ 
 & & & & parameter &  F$_{2-10 {\rm keV}}$ & F$_{1.4 {\rm GHz}}$\\ 
 & & [$10^{20}$\,cm$^{-2}$] & &  &  [erg\,cm$^{-2}$\,s$^{-1}$] & [mJy]\\ 
\hline 
1FGL J0106.7+4853 & PSR J0106+4855 & 11.9 (fixed) & PL & $\Gamma$=2.0 (fixed) & $<$ 1.0 $\times$ 10$^{-14}$ & $<$ 0.008 \\

1FGL J1312.6+0048 & source & 0.0 (fixed) & PL & $\Gamma$=1.9$^{+0.4}_{-0.3}$ & 8.0$^{+2.9}_{-2.6}$ $\times$ 10$^{-14}$ & $<$ 0.71 \\ 
  & off-source$^c$ & 2.1 (fixed) & PL & $\Gamma$=2.0 (fixed) & $<$ 9.3 $\times$ 10$^{-14}$ & $<$ 0.82\\ 
 
1FGL J1739.4+8717 & source & 6.36 (fixed) & PL & $\Gamma$=2.1$^{+0.5}_{-0.4}$ & 3.6 $^{+2.0}_{-1.5}$ $\times$ 10$^{-13}$ & $<$ 0.58\\ 
                  & NVSS J173722+871744 & 6.4 (fixed) & PL & $\Gamma$=2.0 (fixed) & $<$ 1.5 $\times$ 10$^{-14}$ & 61.3 \\ 
 
1FGL J1743.8$-$7620 & - & 8.13 (fixed) & PL & $\Gamma$=2.0 (fixed) & $<$ 4.5 $\times$ 10$^{-14}$ & -\\ 
1FGL J1902.0$-$5110 & PSR J1902-5105 & 4.87 (fixed) & PL & $\Gamma$=2.0 (fixed) & $<$ 2.5 $\times$ 10$^{-14}$ & -\\ 
1FGL J2043.2+1709 & src A & 6.63 (fixed) & PL & $\Gamma$=1.67$^{+0.16}_{-0.15}$ & 2.1$^{+0.4}_{-0.4}$ $\times$ 10$^{-13}$ & $<$ 1.01\\ 
                  & src B & 0.0 (fixed) & PL & $\Gamma$=1.6$^{+0.4}_{-0.3}$ & 9.1$^{+3.5}_{-3.0}$ $\times$ 10$^{-14}$ & $<$ 1.00\\ 
                  & PSR J2043+1711 & 6.67 (fixed) & PL & $\Gamma$=2.0 (fixed) & $<$ 3.6 $\times$ 10$^{-14}$ & 0.01$^{a}$\\ 
1FGL J2302.8+4443 & PSR J2302+4442 & 0.0 (fixed) & BB & $kT$=0.31$^{+0.06}_{-0.05}$\,keV & 3.3$^{+2.7}_{-2.0}$ $\times$ 10$^{-15}$ & 1.2$^{b}$\\ 
\hline 
\end{tabular} 
\end{center} 
  Note. ---Summary of analysis results of the seven studied 1FGL sources. Unabsorbed X-ray flux and 90 $\%$ confidence upper limits 
in the 2$-$10 keV band for all the sources are listed. The X-ray upper limits are calculated assuming a power law model 
with a photon index of 2 for the 1FGL sources with no X-ray detection. In the seventh column, radio fluxes or upper limits at 1.4 GHz 
are presented. Radio fluxes and upper limits 
The radio values with no footnote marks were taken either directly from the NVSS (Condon et al. 1998) catalog (for NVSS J173722+871744) or measured by us from the NVSS images (the upper limits; 90$\%$ confidence levels).\\ 
$^{a}$ This value is estimated by averaging the radio flux over four observations during which the pulsar was detected.\\ 
$^{b}$ From Cognard et al. (2011). \\
$^{c}$ The evaluated off-source X-ray upper limits expected to represent MSP PSR J1312+00.
\label{tab:fits} 
\end{table}

\begin{figure}[m] 
\begin{center} 
\includegraphics[width=150mm]{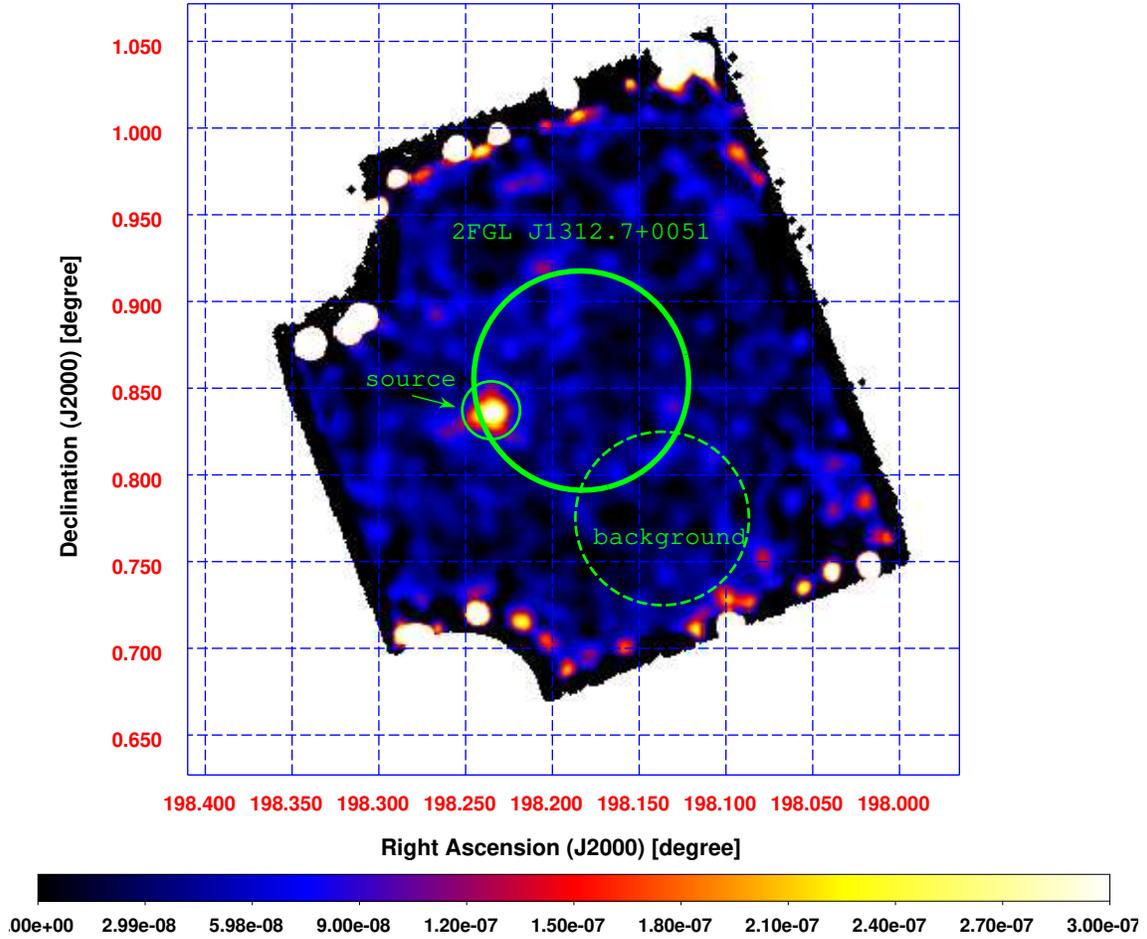} 
\end{center} 
\caption{X-ray image of 1FGL~J1312.6+0048 by {\it Suzaku}/XIS0+3 (FI CCDs) in the 0.5 to 10 keV energy band. Thick solid ellipse denotes the 95\% position error of 1FGL~J1312.6+0048 in the 2FGL catalog. Thin solid and dashed circles show the source and background regions, respectively. The accurate position of PSR J1312+00 is still not available in the literature.} 
\label{fig:J1312_X_image} 
\end{figure} 
 
\begin{figure}[m] 
\begin{center} 
\includegraphics[width=150mm]{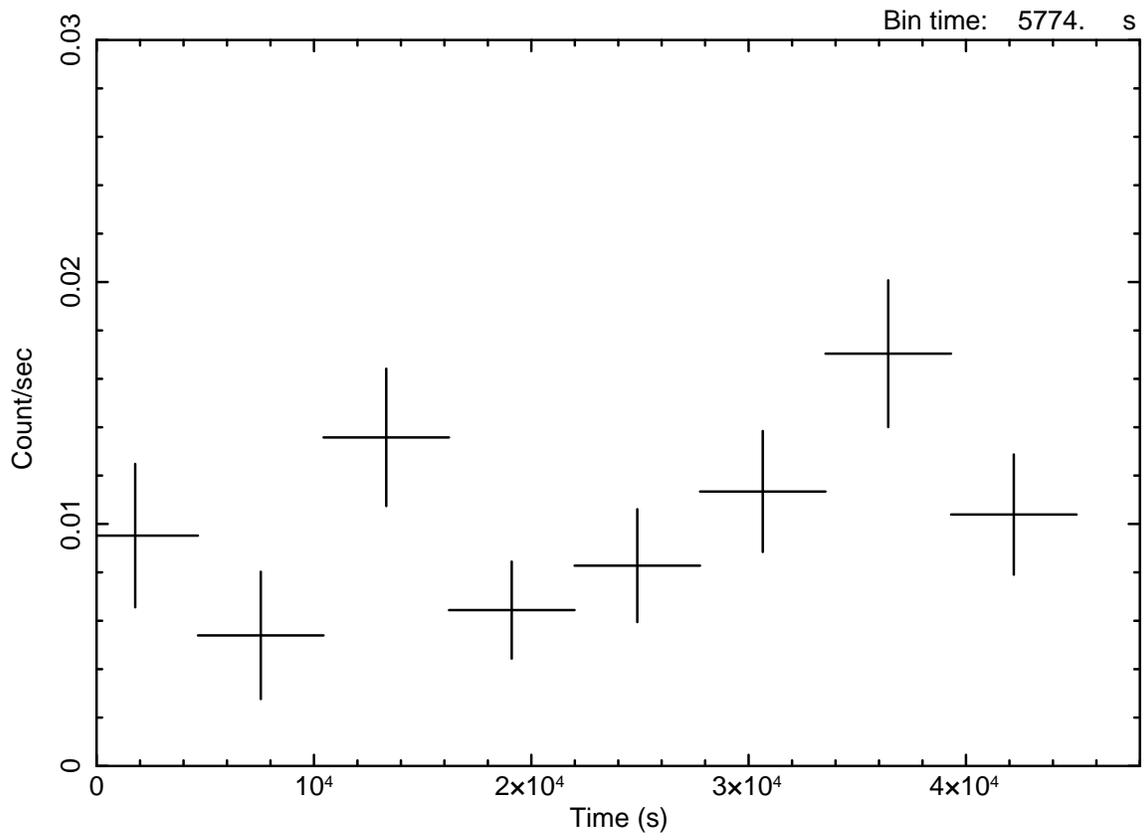} 
\end{center} 
\caption{{\it Suzaku}/XIS light curve of the possible X-ray counterpart of 1FGL~J1312.6+0048.} 
\label{fig:J1312_X_lc} 
\end{figure} 
 
\begin{figure}[m] 
\begin{center} 
\includegraphics[width=150mm]{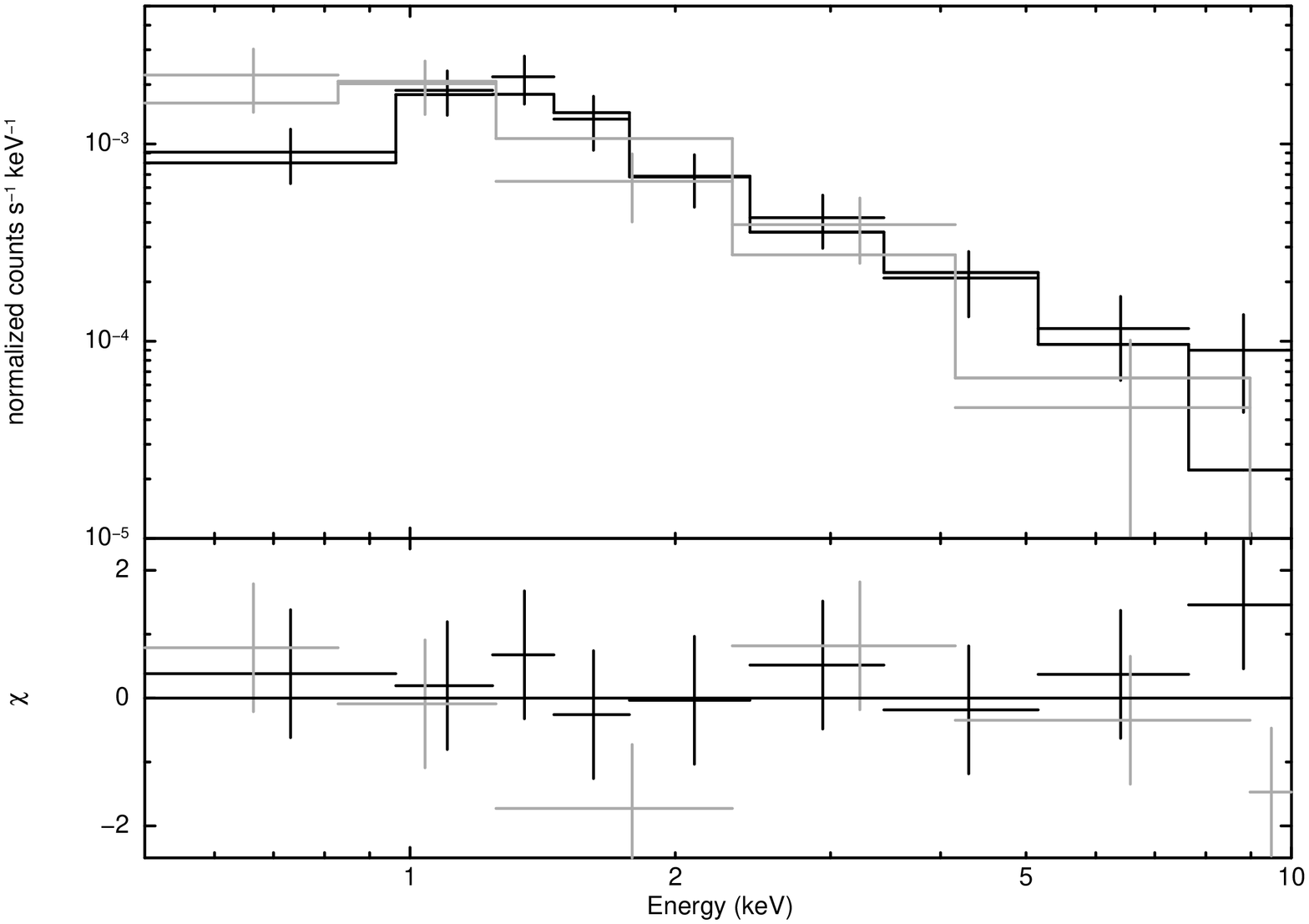} 
\end{center} 
\caption{{\it Suzaku}/XIS spectrum of the possible X-ray counterpart for 1FGL~J1312.6+0048 
fitted with a power-law model. Black plots show the FI data and gray plots show the BI data.} 
\label{fig:J1312_X_spec} 
\end{figure} 

\begin{figure}[m] 
\begin{center} 
\includegraphics[width=150mm]{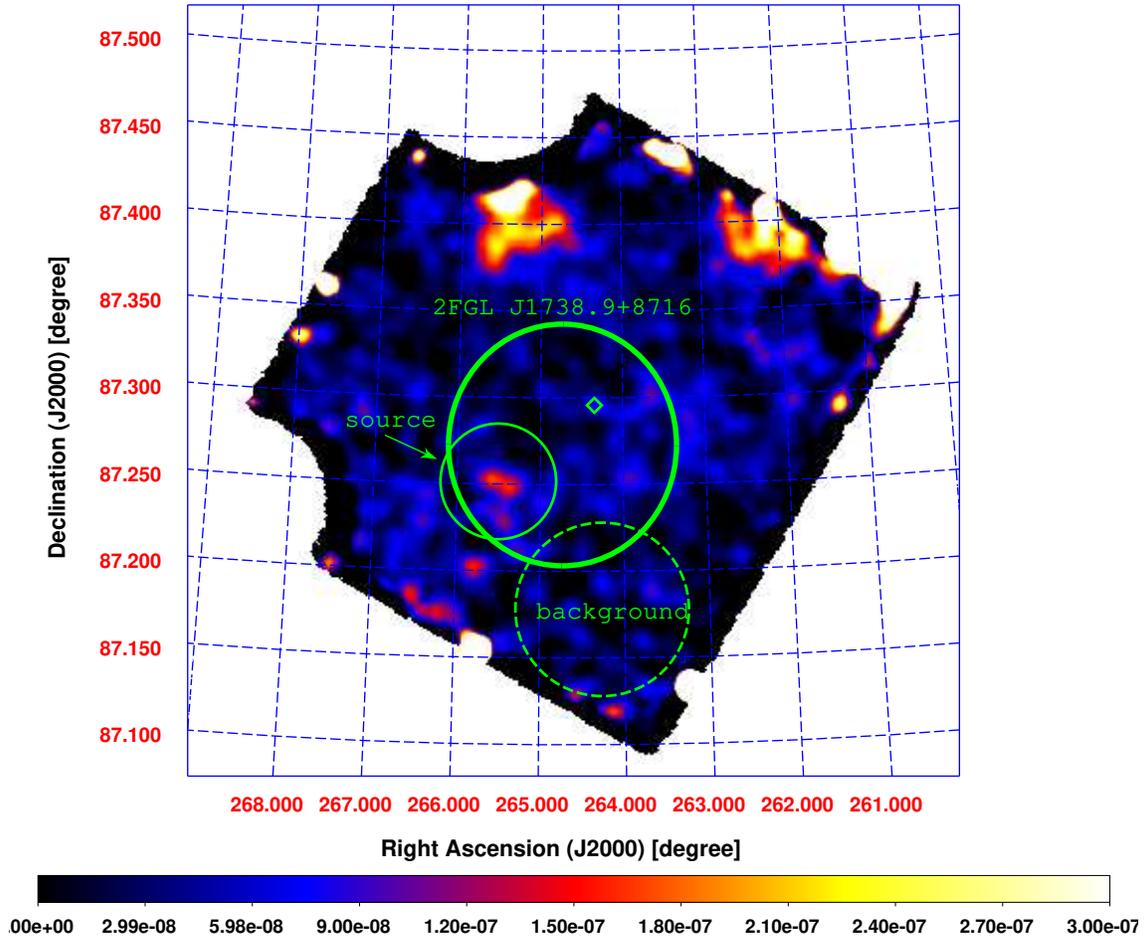} 
\end{center} 
\caption{X-ray image of 1FGL~J1739.4+8717 by {\it Suzaku}/XIS0+3 (FI CCDs) in the 0.5 to 10 keV energy band. Thick solid ellipse denotes the 95\% position error of the 2FGL catalog counterpart of 1FGL~J1739.4+8717. Thin solid and dashed circles show the source and background regions, respectively. The radio position of NVSS J173722+871744 is marked with a green rhombus.} 
\label{fig:J1739_Ximage} 
\end{figure}

\clearpage 
\begin{figure}[m] 
\begin{center} 
\includegraphics[width=150mm]{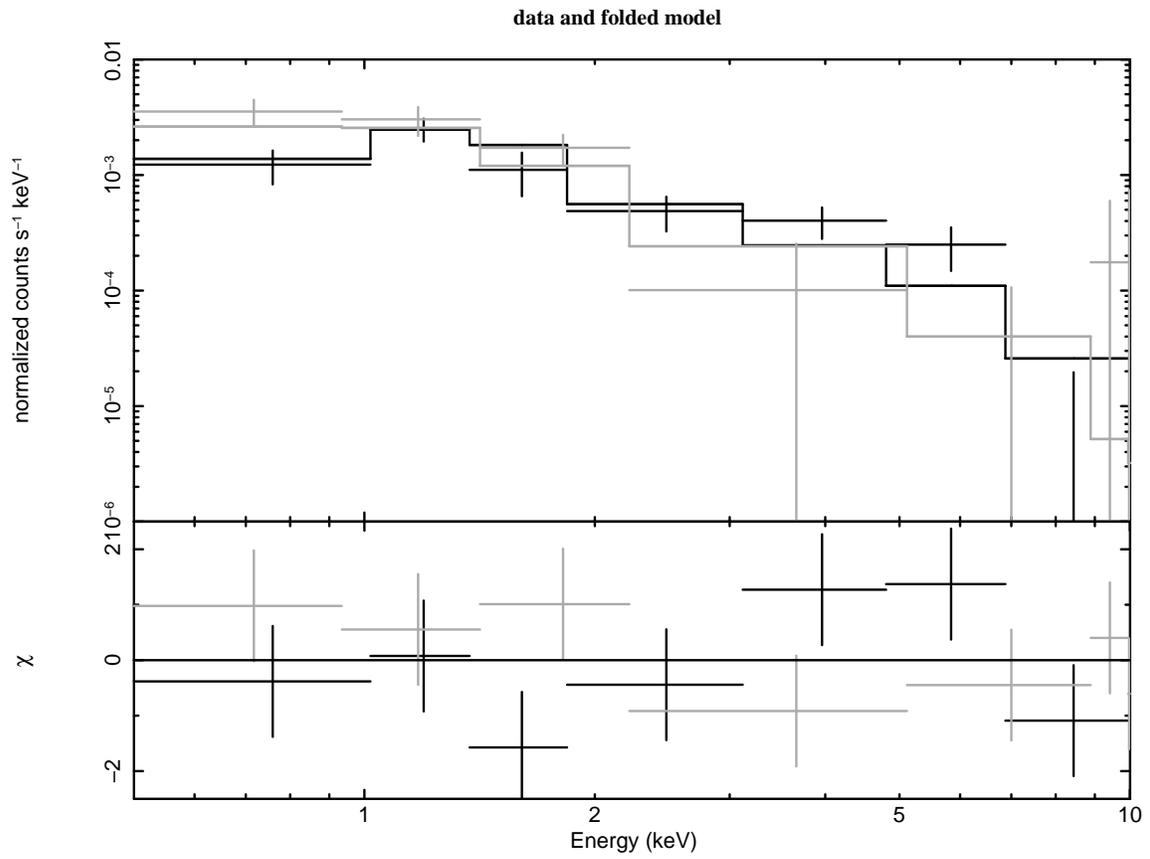} 
\end{center} 
\caption{X-ray spectrum of the X-ray counterpart of 1FGL~J1739.4+8717 fitted with 
a power-law model. Black plots show the FI data and gray plots show the BI data.} 
\label{fig:J1739_xspec} 
\end{figure} 
 
\clearpage 
\begin{figure}[m] 
\begin{center} 
\includegraphics[width=150mm]{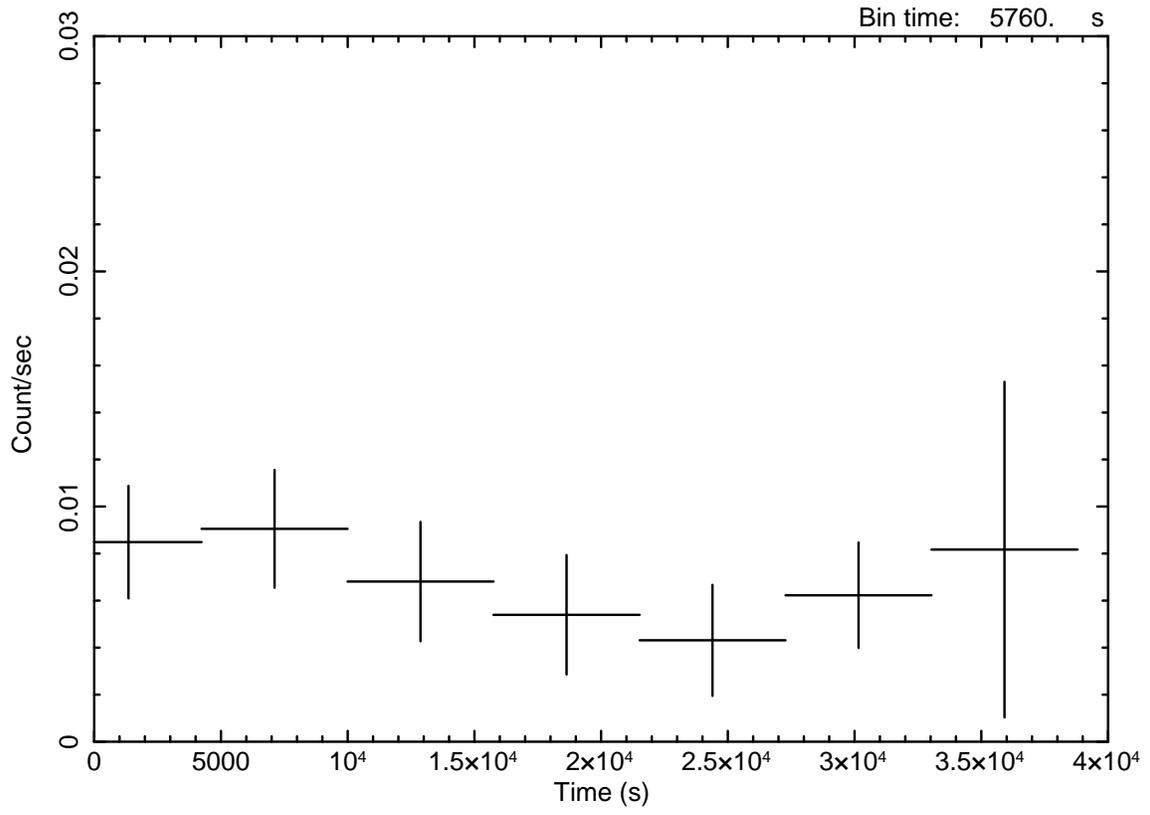} 
\end{center} 
\caption{\SU/XIS light curve of the X-ray counterpart of 1FGL~J1739.4+8717.} 
\label{fig:J1739_lc}
\end{figure} 
 
\clearpage 
\begin{figure}[m] 
\begin{center} 
\includegraphics[width=150mm]{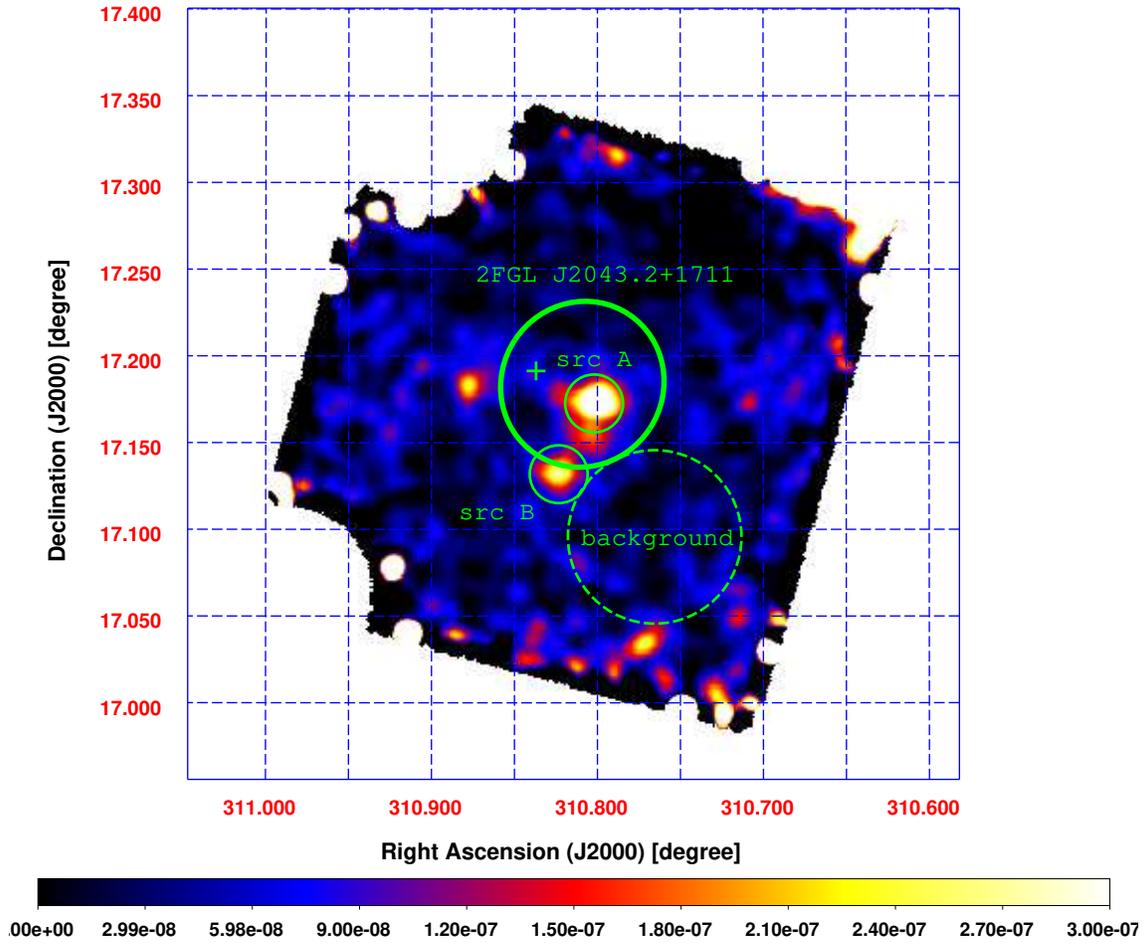} 
\end{center} 
\caption{X-ray image of 1FGL~J2043.2+1709 by {\it Suzaku}/XIS0+3 (FI CCDs) in the 0.5 to 10 keV energy band. The thick solid ellipse denotes the 95$\%$ position error of 1FGL~J2043.2+1709 from the 2FGL catalog. Thin solid and dashed circles show the source and background regions, respectively. Two X-ray sources inside the positional error ellipse of 1FGL~J2043.2+1709 are named srcA for the northern source and srcB for the southern source. The position of the associated MSP PSR~J2043+1711 is marked with a cross.} 
\label{fig:J2043_image} 
\end{figure} 
 
\clearpage 
 
\begin{figure}[m] 
  \begin{center} 
    \begin{minipage}{.70\linewidth} 
      \subfigure[Source A]{\includegraphics[width=120mm]{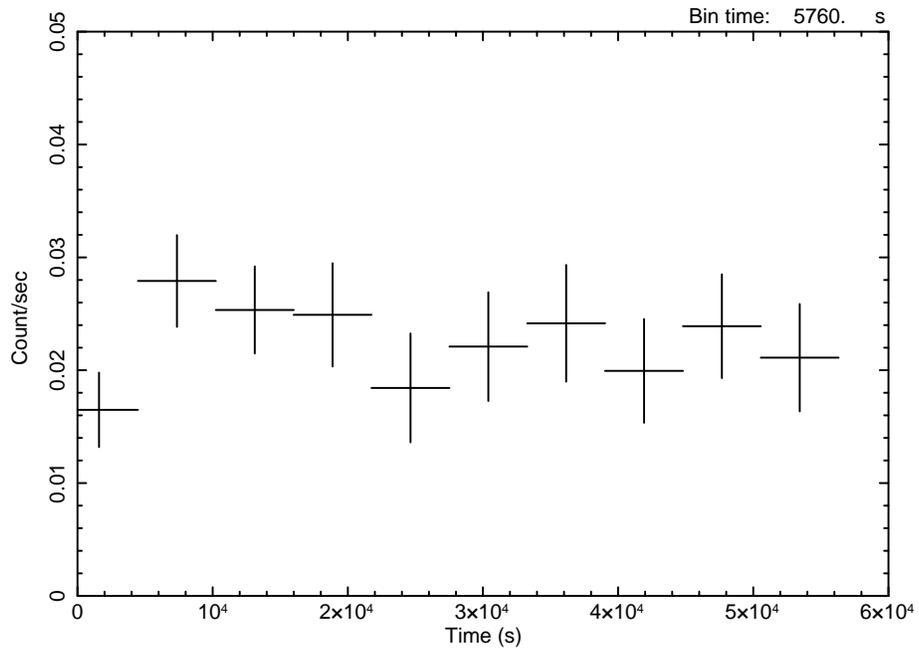}} 
    \end{minipage} 
  \end{center} 
  \begin{center} 
    \begin{minipage}{.70\linewidth} 
      \subfigure[Source B]{\includegraphics[width=120mm]{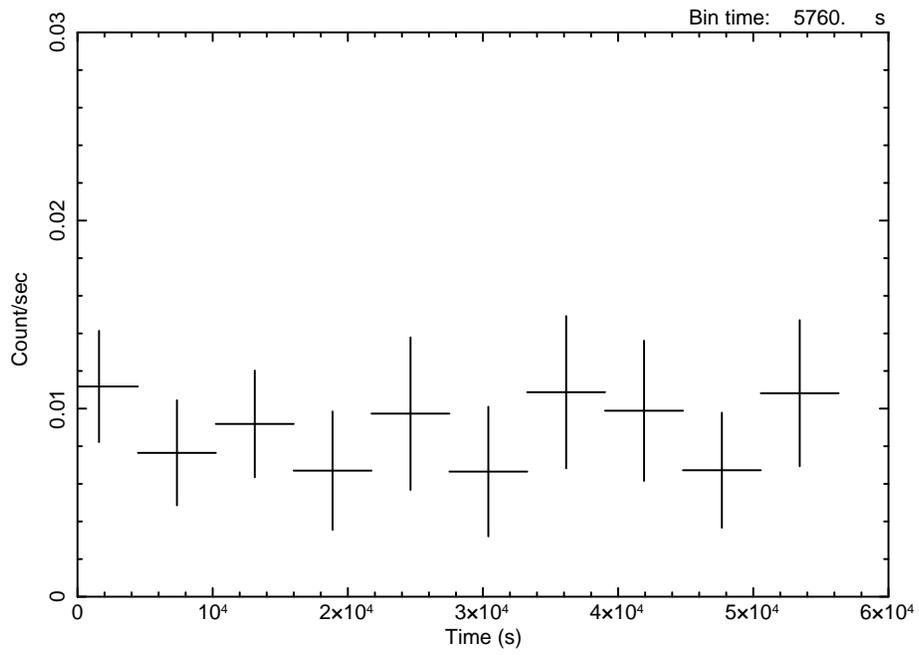}} 
    \end{minipage} 
    \caption{{\it Suzaku}/XIS light curve of srcA (a) and srcB (b).} 
    \label{fig:J2043_X_lc} 
  \end{center} 
\end{figure} 
 
\begin{figure}[m] 
  \begin{center} 
    \begin{minipage}{.70\linewidth} 
      \subfigure[Source A]{\includegraphics[width=120mm]{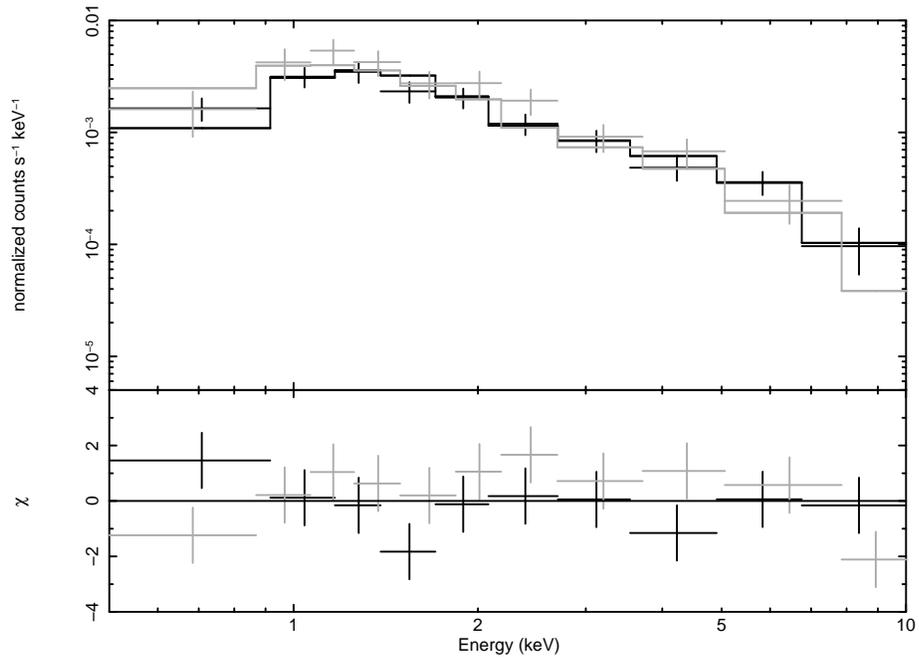}} 
    \end{minipage} 
  \end{center} 
  \begin{center} 
    \begin{minipage}{.70\linewidth} 
      \subfigure[Source B]{\includegraphics[width=120mm]{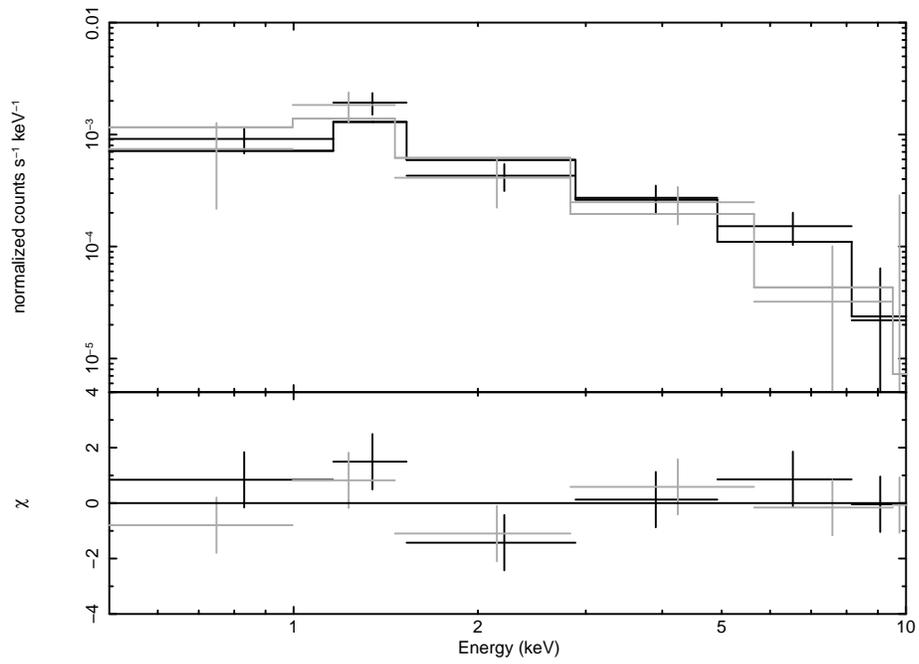}} 
    \end{minipage} 
    \caption{{\it Suzaku}/XIS spectra of srcA (a) and srcB (b) inside the  
error ellipse of 1FGL~J2043.2+1709. Both spectra are fitted with an absorbed  
power-law model. Black plots show the FI data and gray plots show the BI data.} 
    \label{fig:J2043_spectrum} 
  \end{center}
\end{figure}

\clearpage 
\begin{figure}[m] 
\begin{center} 
\includegraphics[width=150mm]{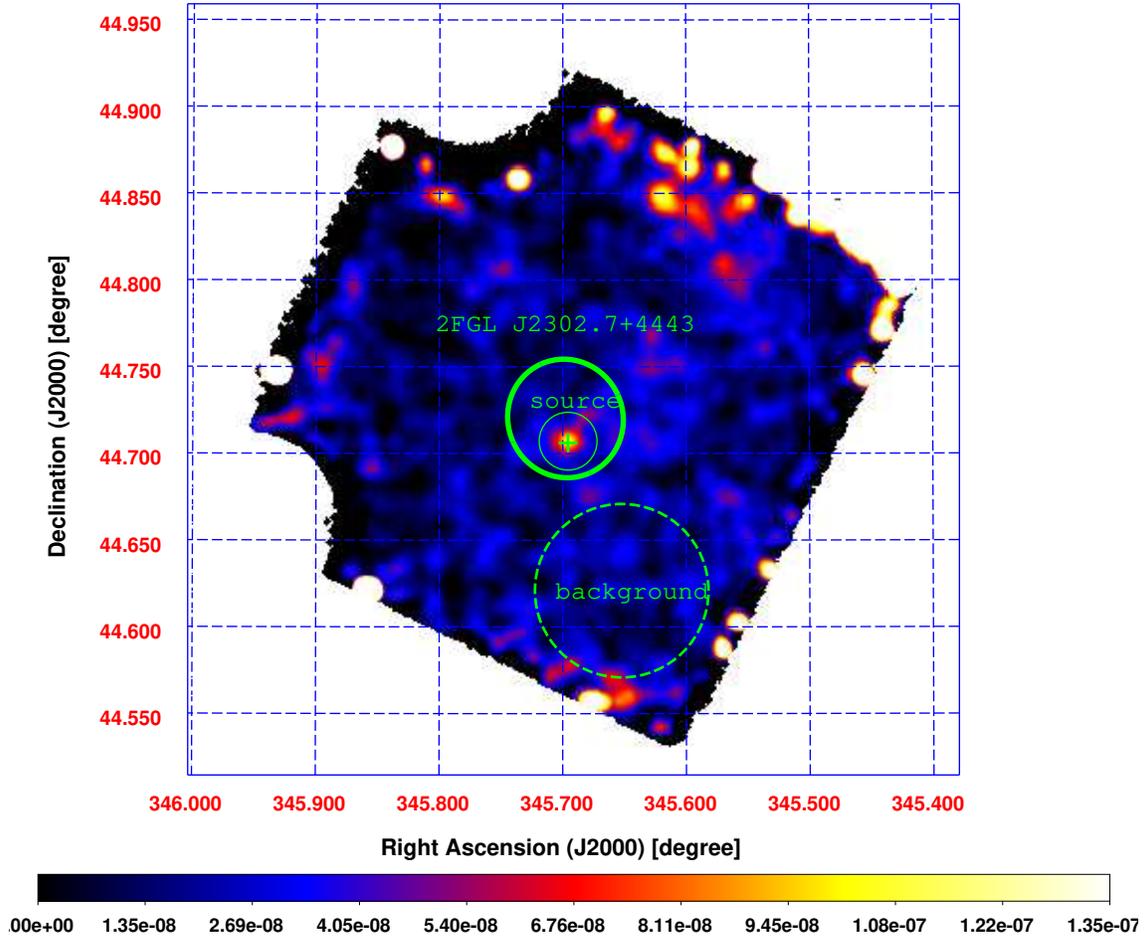}
\end{center}
\caption{X-ray image of 1FGL~J2302.8+4443 by {\it Suzaku}/XIS0+3 (FI CCDs) in the 0.5 to 2 keV energy band. Thick solid ellipse denotes 95$\%$ position error of 1FGL~J2302.8+4443 from the 2FGL catalog. Thin solid and dashed circles show the source and background regions, respectively. The position of the associated MSP PSR~J2302+4442 is marked with a cross.} 
\label{fig:J2302_image} 
\end{figure} 
 
\clearpage 
\begin{figure}[m] 
\begin{center} 
\includegraphics[width=150mm]{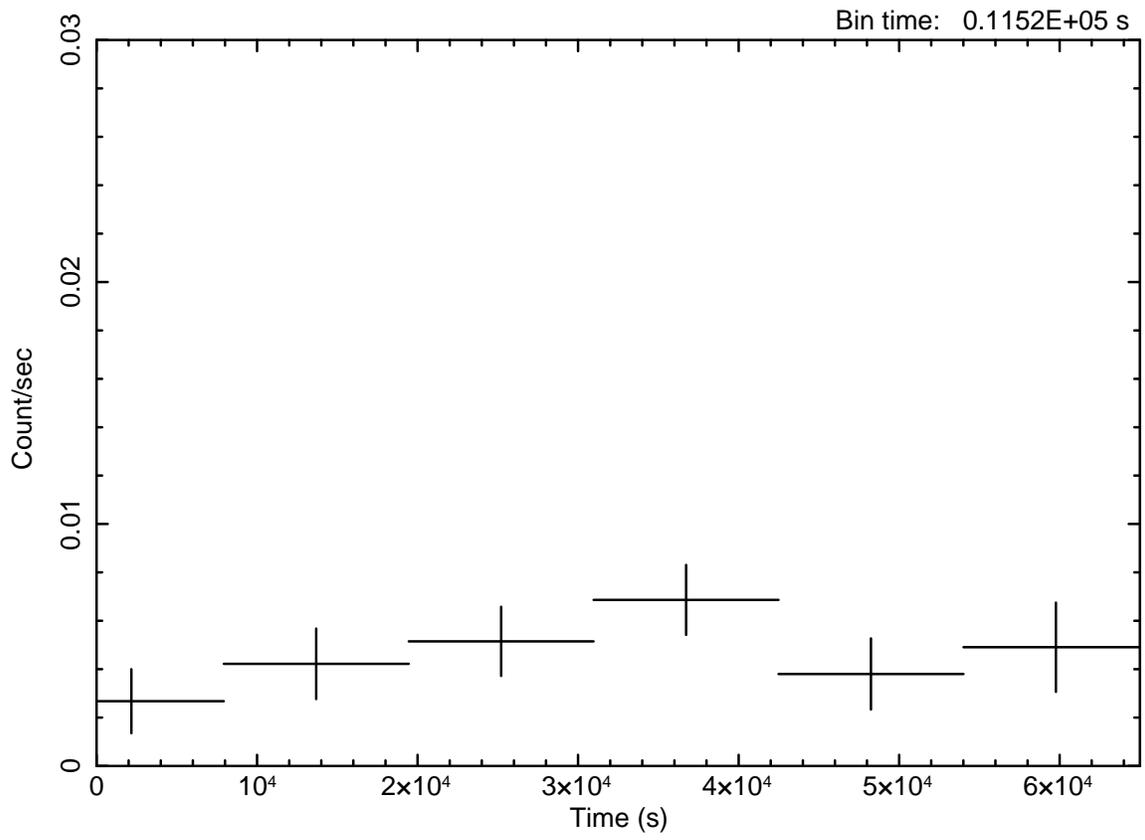} 
\end{center} 
\caption{{\it Suzaku}/XIS light curve of the X-ray counterpart of 1FGL~J2302.8+4443.} 
\label{fig:J2302_X_lc} 
\end{figure}

\clearpage 
\begin{figure}[m] 
\begin{center} 
\includegraphics[width=150mm]{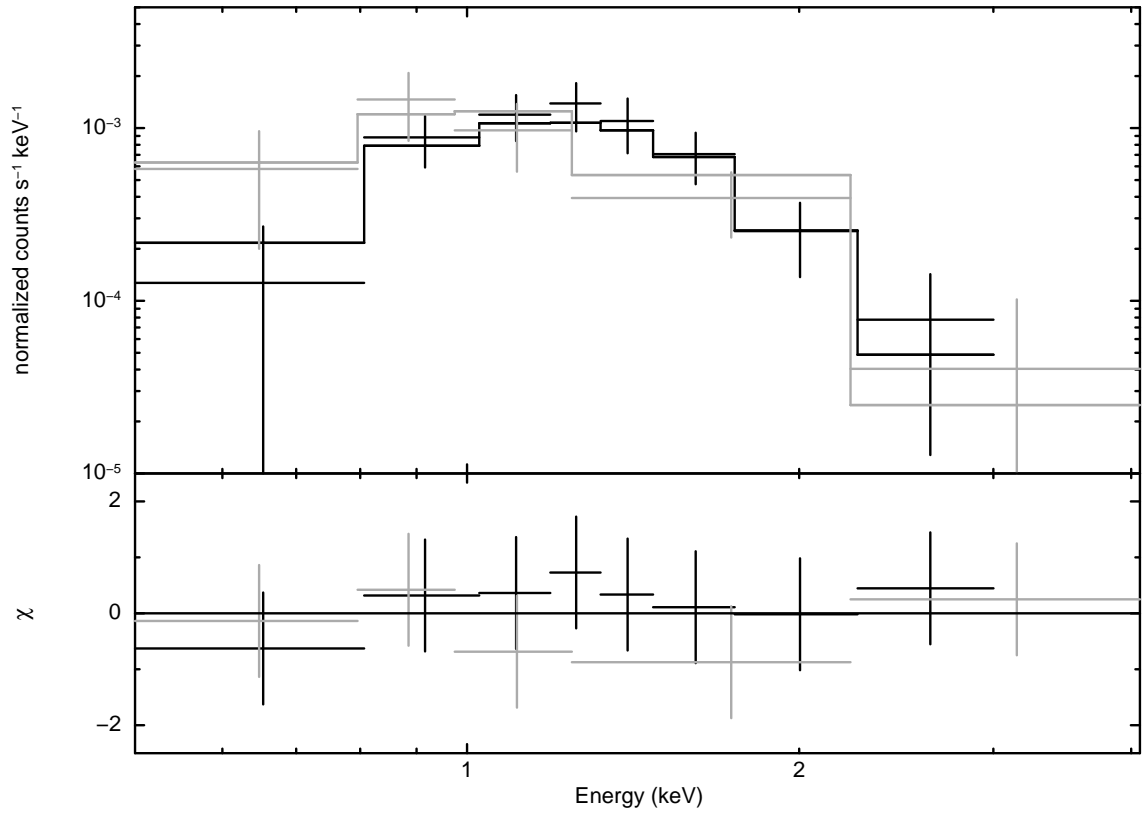} 
\end{center} 
\caption{{\it Suzaku}/XIS spectrum of the X-ray counterpart of 1FGL~J2302.8+4443 fitted with 
a black body model. Black plots show the FI data and gray plots show the BI data.} 
\label{fig:J2302_Xspec} 
\end{figure}

\begin{figure}[m] 
\begin{center} 
\includegraphics[width=150mm]{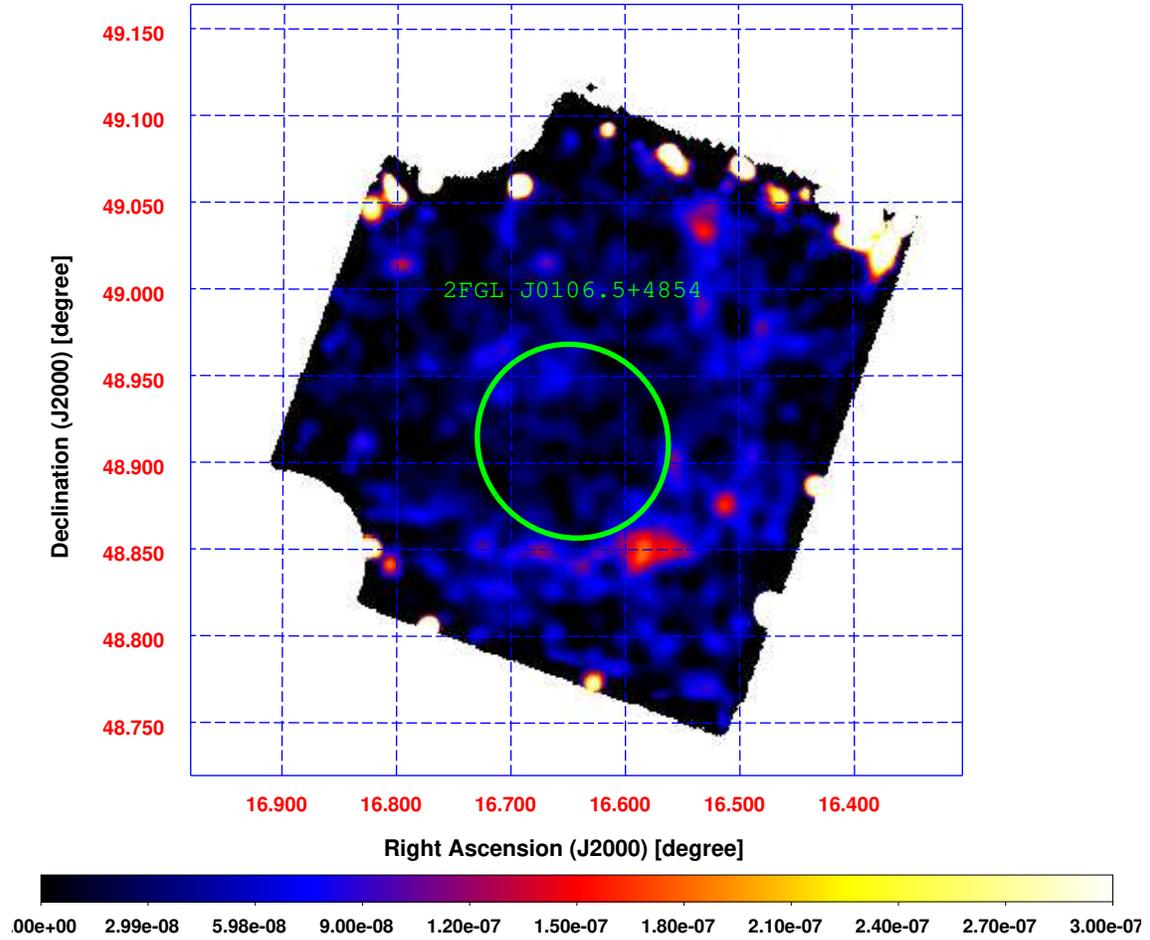} 
\end{center} 
\caption{X-ray image of 1FGL~J0106.7+4853 by {\it Suzaku}/XIS0+3 (FI CCDs) in the 0.5 to 10 keV energy band. Thick solid ellipse denotes 95\% position error of 1FGL~J0106.7+4853 from the 2FGL catalog. } 
\label{fig:J0106_X_image} 
\end{figure} 
 
\begin{figure}[m] 
\begin{center} 
\includegraphics[width=150mm]{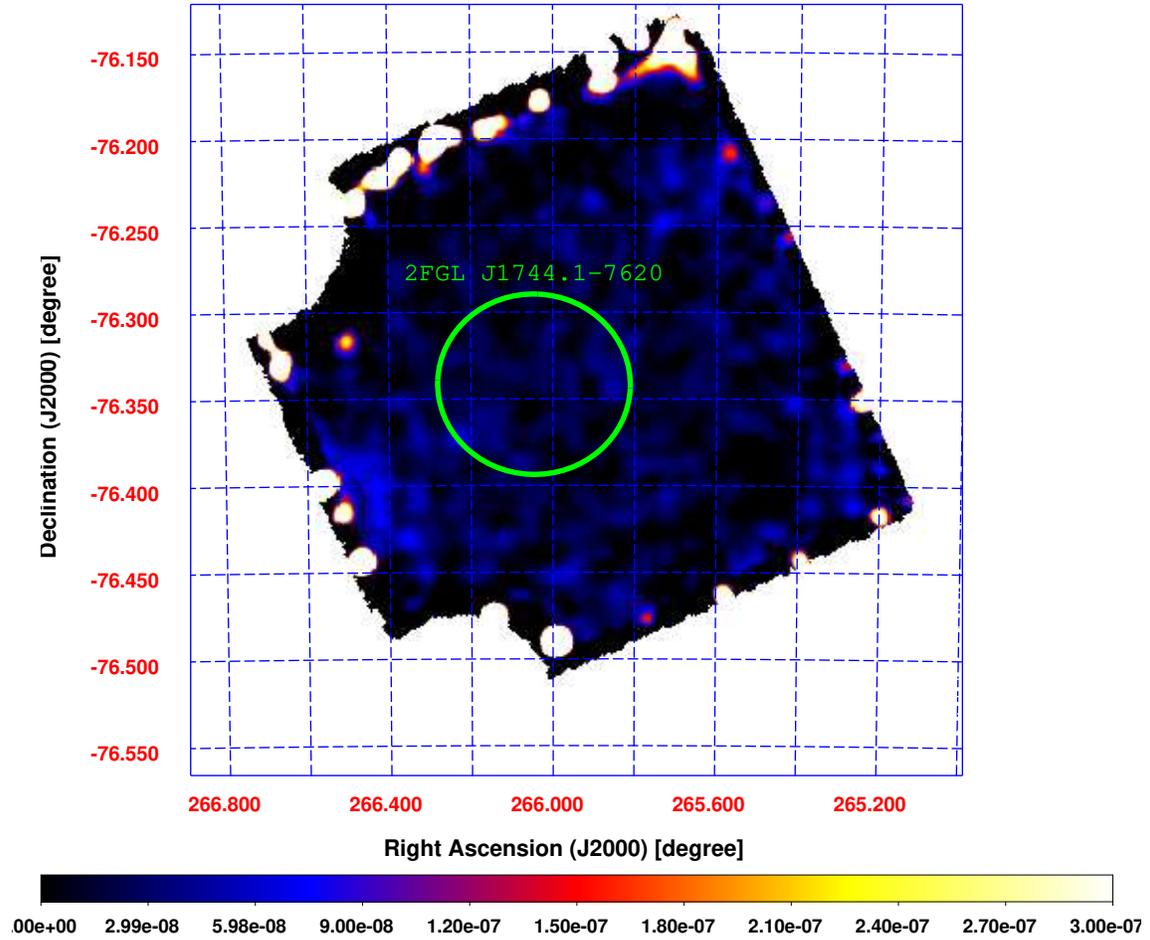} 
\end{center} 
\caption{X-ray image of 1FGL~J1743.8$-$7620 by {\it Suzaku}/XIS0+3 (FI CCDs) in the 0.5 to 10 keV energy band. Thick solid ellipse denotes 95\% position error of 1FGL~J1743.8$-$7620 from the 2FGL catalog.} 
\label{fig:J1743_X_image} 
\end{figure} 
 
\begin{figure}[m] 
\begin{center} 
\includegraphics[width=150mm]{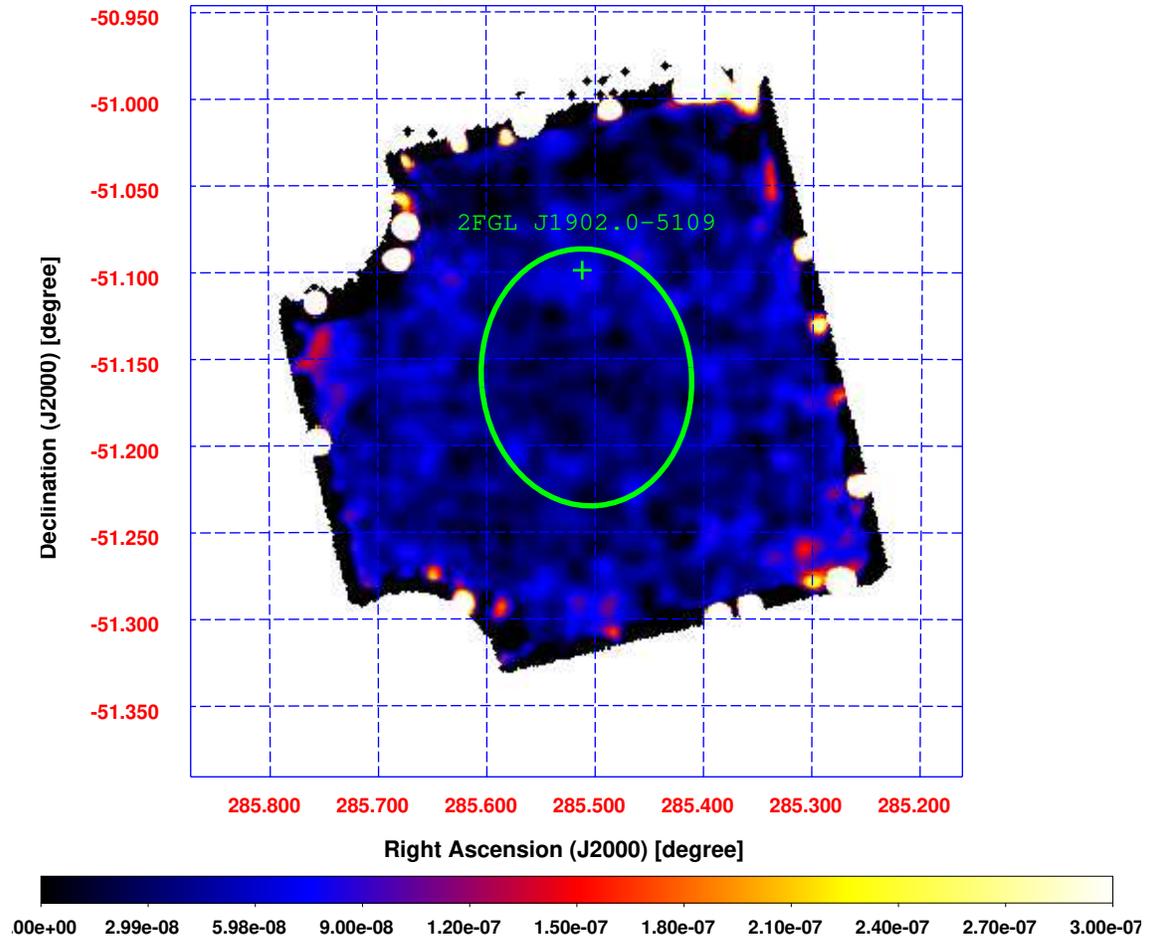} 
\end{center} 
\caption{X-ray image of 1FGL~J1902.0$-$5110 by {\it Suzaku}/XIS0+3 (FI CCDs) in the 0.5 to 10 keV energy band. Thick solid ellipse denotes 95\% position error of 1FGL~J1902.0$-$5110 from the 2FGL catalog. The accurate position of PSR J1902$-$5105 is still not available in the literature.} 
\label{fig:J1902_X_image} 
\end{figure}

\begin{figure}[m] 
\begin{center} 
\includegraphics[width=150mm]{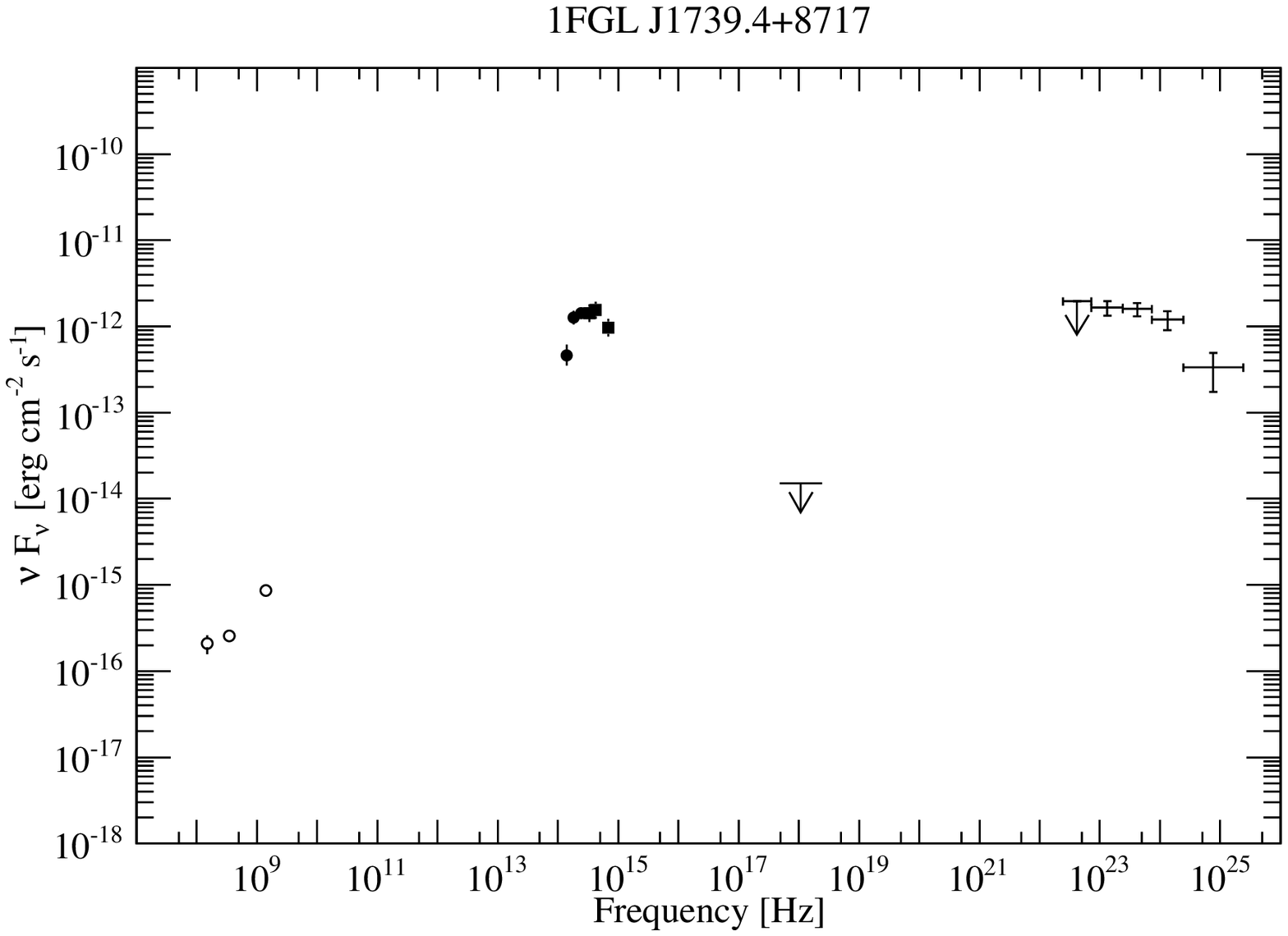} 
\end{center} 
\caption{Spectral energy distributions of 1FGL J1739.4+8717. The X-ray data represents $90\%$ confidence upper limit calculated from our \SU data. The $\gamma$-ray data points are taken from the 1FGL catalog \citep{1FGL}. The radio data points are 6C B175708+871924 in 6C catalog \citep{bal85}, WENSS B1758.4+8718 in WENSS catalog \citep{ren97} and NVSS J173722+871744 in NVSS catalog \citep{con98}. Infrared and optical plots, 2MASS J17372480+8717433 and USNOB 1772-0020476, are quoted from 2MASS point source catalog \citep{2MASS} and USNO B1.0 catalog \citep{usb1}, respectively.}
\label{fig:J1739sed} 
\end{figure}

\begin{figure}[m] 
\begin{center} 
\includegraphics[width=180mm]{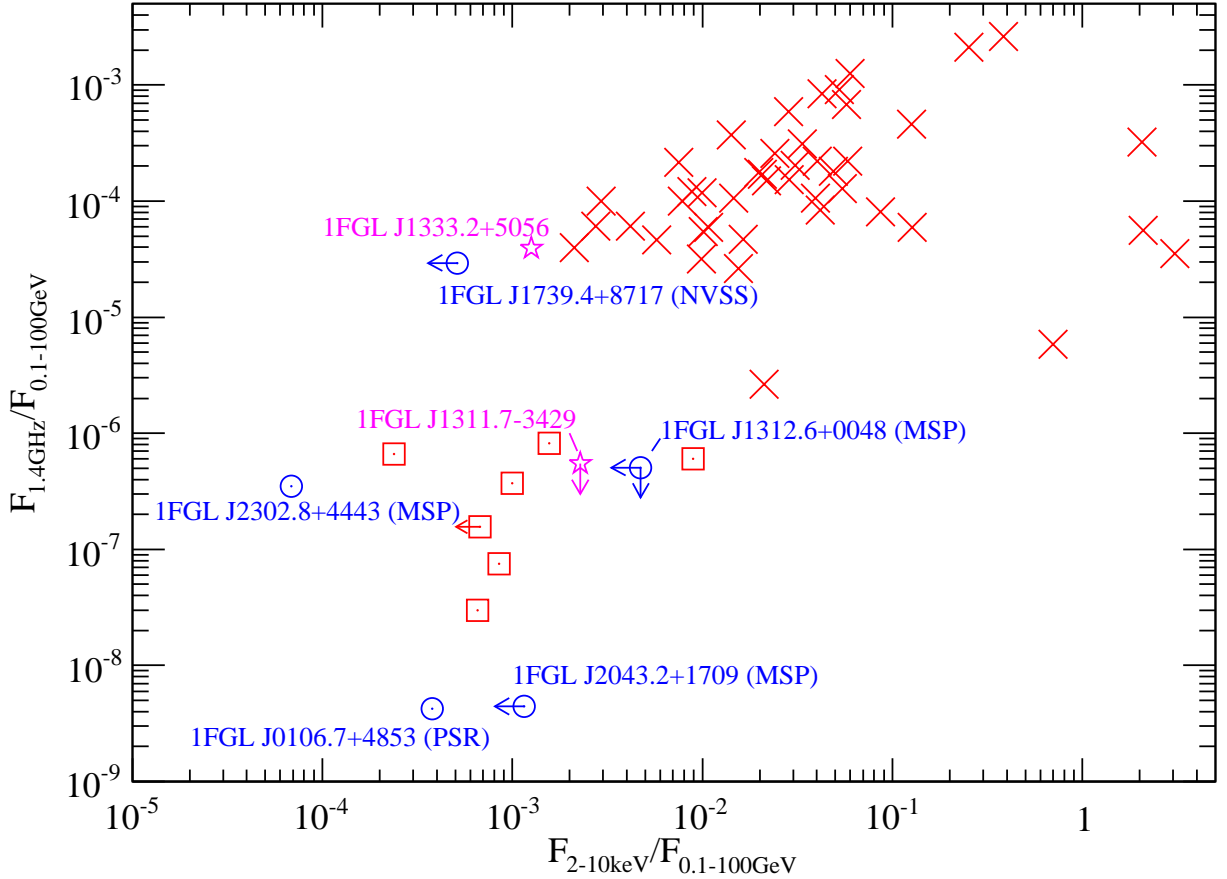} 
\end{center} 
\caption{X-ray to $\gamma$-ray flux ratios versus radio to $\gamma$-ray flux ratios for 
unassociated \F sources. The blue circles show the  
flux ratios of unassociated \F sources which 
we observed using the {\it Suzaku} satellite. The $\gamma$-ray flux and the radio upper  
limits of these sources are taken from values of the 1FGL catalog  
and the NVSS catalog (Condon et al. 1998),  
respectively. Red x marks indicate the data plots of blazars listed in  
Abdo et al. (2010f). 
Red squares represent the data points of the \F sources that are  
associated with MSPs discovered by Cognard et al. (2011) and 
Ransom et al. (2011), and three \F sources listed in Abdo et al. (2010j) for which 
the best fit model of the X-ray spectra were determined. The values of 
X-ray flux and radio fluxes of MSPs in Abdo et al. (2010j) are quoted and  
derived from Bailes et al. (1997), Lundgen et al. (1995), Navarro et al. (1995), 
Sakurai et al. (2001), Webb et al. (2004a), and Webb et al. (2004b). For the MSPs we cannot quote the
radio flux at 1.4 GHz, we estimated 1.4 GHz flux by the extrapolation using radio fluxes 
that are available in the literature assuming a power-law model.
Pink stars show the two still unassociated 1FGL sources from our first year campaign targets (Maeda et al. 2011) and another two 1FGL sources studied in Maeda et al. (2011) and that are now associated with MSPs are involved in red squares.} 
\label{fig:ratio} 
\end{figure}

\end{document}